\documentclass[pdflatex,sn-mathphys-ay]{sn-jnl}

\usepackage{xurl}
\usepackage{hyperref}
\usepackage{graphicx}%
\usepackage{multirow}%
\usepackage{amsmath,amssymb,amsfonts}%
\usepackage{amsthm}%
\usepackage{mathrsfs}%
\usepackage[title]{appendix}
\usepackage{xcolor}%
\usepackage{textcomp}%
\usepackage{manyfoot}%
\usepackage{booktabs}%
\usepackage{algorithm}%
\usepackage{algorithmicx}%
\usepackage{algpseudocode}%
\usepackage{listings}%
\usepackage{natbib}
\usepackage{rotating}
\usepackage{csquotes} 
\usepackage{float} 
\usepackage[most]{tcolorbox} 
\usepackage{adjustbox}  

\floatstyle{plain}
\newfloat{quotebox}{htbp}{loq}
\floatname{quotebox}{Quote}

\begin{document}

\title[Article Title]{Investigating the originality of scientific papers across time and domain: A quantitative analysis}

\author[1, *]{\fnm{Jack H.} \sur{Culbert}\,\href{https://orcid.org/0009-0000-1581-4021}{\orcid{}}}
\author[2]{\fnm{Yoed N.} \sur{Kenett}\,\href{https://orcid.org/0000-0003-3872-7689}{\orcid{}}}
\author[1]{\fnm{Philipp} \sur{Mayr}\,\href{https://orcid.org/0000-0002-6656-1658}{\orcid{}}}

\date{December 2025}

\affil[1]{\orgname{GESIS -- Leibniz Institute for the Social Sciences}, 
\country{Germany}}

\affil[2]{\orgdiv{Faculty of Data and Decision Sciences}, \orgname{Technion Israel Institute of Technology}, 
\city{Haifa}, 
\country{Israel}}
\affil[*]{Corresponding Author: Jack H. Culbert, \href{mailto:jack.culbert@gesis.org}{jack.culbert@gesis.org}}

\abstract{The study of creativity in science has long sought quantitative metrics capable of capturing the originality of the scientific insights contained within articles and other scientific works. In recent years, the field has witnessed a substantial expansion of research activity, enabled by advances in natural language processing and network analysis, and has utilised both macro- and micro-scale approaches with success. However, they often do not examine the text itself for evidence of originality. In this paper, we apply a computational measure correlating with originality from creativity science, Divergent Semantic Integration (DSI), to a set of 51,200 scientific abstracts and titles sourced from the Web of Science. To adapt DSI for application to scientific texts, we advance the original BERT method by incorporating SciBERT—a model trained on scientific corpora—into the computation of DSI. 
In our study, we observe that DSI plays a more pronounced role in the accrual of early citations for papers with fewer authors, varies substantially across subjects and research fields, and exhibits a declining correlation with citation counts over time. 
Furthermore, by modelling SciBERT- and BERT-DSI as predictors of the logarithm of 5-year citation counts—alongside field, publication year, and the logarithm of author count—we find statistically significant relationships, with adjusted R$^{2}$ of 0.103 and 0.101 for BERT-DSI and SciBERT-DSI. Because existing scientometric measures rarely assess the originality expressed in textual content, DSI provides a valuable means of directly quantifying the conceptual originality embedded in scientific writing.}

\keywords{Creativity, Scientific Novelty, Scientific Originality, Transformer-based language models, Contextualized word embeddings, Scientometrics}

\maketitle

\section{Introduction}
One aspect of abstracts that varies across scientific domains—and evolves over time—is their degree of originality. While some scientific domains have strict norms on abstract formats and content, the increased challenge of a scientific paper getting attention, due to the rapid increase in volume of papers with decreased attention span due to information overload \citep{holyst024}, likely impacts the originality of abstracts. 
However, the impact of such pressures on abstract writing and title formulation could have both a facilitative or inhibitory impact on their originality: Abstracts and titles may become more original over time, to compete for a reader’s attention more strongly, or they may become less original, to standardize within scientific disciplines and minimize information overload. A possible way to examine these competing hypotheses is by harnessing computational tools that have been recently developed in creativity research to quantitatively assess the originality of short narratives. Specifically, an approach called Divergent Semantic Integration \citep{Johnson2023, Patterson2025-wb}.

Creativity is defined as the ability to generate ideas that are both original and useful \citep{Green02072024,Runco2012-ou}. Over the past decade, the introduction of computational metrics in creativity research has propelled the field forward, increasingly introducing new metrics to provide quantitative and objective measures of this complex capacity \citep{Kenett2019-ub, Kenett2024-jv, Beaty2021-bu, ORGANISCIAK2023101356,Patterson2025-wb}. These metrics capitalize on advancements in natural language processing, and have especially focused on quantifying idea originality \citep{Kenett2019-ub}.

Overall, computational metrics of originality are based on the notion of semantic distance - the conceptual, or semantic dissimilarity between concepts in a language model, such as Word2Vec \citep{Mikolov2013-en}, or NERT \citep{Jiang2023-uk}. Previous research has shown how semantic distance is strongly correlated with subjective originality ratings across varied creativity tasks \citep{Beaty2021-bu, Gray2019-zf, Campidelli2026-sb}. 

\subsection{Divergent Semantic Integration}

Divergent Semantic Integration (DSI) \citep{Johnson2023} is a computational metric for short textual narratives, which was shown to correlate with subjective ratings of originality. DSI is computed as the arithmetic mean of cosine distances between embeddings of sentences from a language model, measuring the overall richness of the language used by the writer in their narrative. 

The underlying idea of DSI is that divergent concepts within a text map to distant regions of the model’s embedding space; consequently, more diverse concepts are, on average, farther apart than similar or conventional ones, yielding a higher DSI score. Thus, higher DSI scores indicate a richer, more diverse and original content of the narrative analysed. Extensive empirical research in creativity has shown that highly creative individuals possess richer memory structures and are able to search, combine, and expand ideas more broadly, enabling the generation of original insights \citep{Beaty2023, Benedek2023}.

Since the release of DSI, it has been studied primarily in the context of linguistic creativity \citep{11126815}, but also a particular focus in measuring the creativity of large language models and comparing them to human creativity \citep{orwig2024}. Significant work has been done in automating measurement of scientific creativity or originality in scientific texts, described here in Section \ref{sec:creativity}. However, to the best of our knowledge, DSI was first applied to scientific titles and abstracts in our previous study \citep{culbertISSI2025}. 

\subsection{Creativity in Science}\label{sec:creativity}

This study builds upon previous research into creativity in science, also frequently referred to as originality, scientific innovation, breakthroughs, or novelty. \cite{Zhao2025-uz} categorizes scientific originality into four dimensions: uniqueness, or the discovery of new knowledge; recombination, or the recombination of knowledge elements; bridging, or new links connecting two knowledge clusters; and surprise, or the subjective cognitive violation of expectations. 
\cite{Zhao2025-uz} classifies measures of originality in academic papers into three main categories based on the data types used—citation relations, textual data, and multi-type approaches—and further subdivides these into the following specific originality measures. They include: reference-based, keyword/entity-level, sentence-level, text-and-citation-relations-based, and text-and-network-structure-based measures. Our approach does not neatly fit into any of the above, as it directly exploits the text itself without extraction techniques as in keyword-entity level measures.

Referenced based approaches such as \cite{TRAPIDO20151488}, \cite{Shibayama2020}, \cite{WANG20171416} and \cite{doi:10.1126/science.1240474}, keyword and entity based approaches such as \cite{https://doi.org/10.1111/j.1756-2171.2011.00140.x}, \cite{LUO2022101282}, \cite{Boudreau2016}, and \cite{doi:10.1177/01655515231161133}, and multi-type approaches such as \cite{shibayama2021} utilise the metadata of a paper to identify novel combinations, unexpected or unlikely combinations, outliers, age and frequency distributions, and connectedness and overlap to measure the originality of a paper.

More recent research utilising the availability of sentence-level and entity based bibliometric information from sources such as arXiv\footnote{\url{https://arxiv.org/}}, bioRxiv\footnote{\url{https://www.biorxiv.org/}} and OpenAlex \citep{priem2022openalexfullyopenindexscholarly} has begun appearing. 

Entity based measures are based on extraction techniques such as in \cite{Liu2022-vt} where bio-entities are extracted from COVID-19 papers in which the number of novel bio-entity pairs is comparied to the potential number of bio-entity pairs is used as a measure of originality, or sentence based measures such as \cite{chen2019} utilising n-grams and searching Scopus for originality, \cite{JEON2023101450} using a local outlier factor applied to word embeddings of titles, or \cite{WANG2024101587} utilising BERTopic to extract core knowledge from a paper and the cloud model from fuzzy mathematics to measure originality.

These approaches all relate more closely to the text than metadata-based methods, which Zhao classifies as micro and macro approaches. In the current study, we focus on what would be classified as a micro approach when applying DSI. Critically, we address a research gap in the development of a metric which utilises the text itself through language models to provide a quantitative measure of the paper's originality based on its title and abstract.

Similar to our current study, \cite{shibayama2021} uses the semantic distance of references to estimate scientific originality in a paper. They do so by: studying the cosine distance between all pairs of embeddings of all titles of referenced papers in a given paper using scispaCy \citep{neumann-etal-2019-scispacy}, using a model focused on biomedical natural language processing, as an embedding model, and adopting the q-percentile of these as a measure for a paper's originality. Similarly to our study, \cite{https://doi.org/10.1002/jocb.658} aimed to score scientific originality using a language model (XLM-RoBERTa) on texts, however they focus on both a different modality of text (a scientific creative thinking task, selected from a larger multivariate study in \cite{WOS:000340692800018}), trained the model on their data (in contrast to our untrained approach) and focused on German texts in particular.

Unlike these previous studies, our large-scale study introduces the first use of DSI to scientometrics, and employs (in part) the text of the article itself through embeddings without extraction methods. While we utilise semantic distance in our metric (similar to \cite{shibayama2021} and \cite{LUO2022101282} - who uses the inverse, semantic similarity), we differ from \cite{shibayama2021} through studying the title and abstract of the paper itself, rather than computing the semantic distance between the titles of referenced papers. Critically, our approach allows conducting large-scale scientometric studies, quantitatively capturing a unique aspect of scientific papers that has hardly been studied before--i.e., originality--in relation to standard bibliometrics. All of this is possible through the use of recent widely available language models.

\subsection{The Significance of Embedding Models}\label{sec:significance}
Word embeddings allow a representation of text to encode and utilise the semantic, logical, and cultural meanings within a text. Prior to the current epoch of transformer models, models such as recurrent neural networks \citep{Rumelhart1986LearningIR, osti_6910294}, long short-term memory \citep{10.1162/neco.1997.9.8.1735} and gated recurrent networks \citep{chung2014empiricalevaluationgatedrecurrent} were studied. Following the publishing of \cite{vaswani2023attentionneed} in 2017 an explosion of research into similar models was conducted, an early frontrunner in transformer models was BERT \citep{devlin2019bertpretrainingdeepbidirectional}, which has been used successfully as a general purpose natural processing model for many applications and scientific endeavours.

Another model widely studied is SciBERT \citep{beltagy-etal-2019-scibert}, which is a model created by AllenAI which uses the same model architecture as BERT but instead was trained on a corpus of scientific papers rather than BERT's training corpus of general purpose text: a combination of BooksCorpus \citep{10.1109/ICCV.2015.11} (800 million words) and English Wikipedia (where only text passages were extracted, 2,500 million words). The corpus of texts for SciBERT's training was a random sample of 1.14 million papers from Semantic Scholar, primarily from the biomedical domain and a minority from the computer science domain, this led to a corpus of roughly the same size (3.17 billion tokens) as the corpus on which BERT was originally trained (3.3 billion tokens).

Importantly, in psychological and  cognitive research, language models, such as BERT have propelled forward computational research regarding apsects related to language, memory, and meaning \citep{MANDERA201757,doi:10.1177/1745691619861372,doi:10.1080/17470218.2014.988735}. With regard to scientific texts, both BERT and SciBERT have proven themselves competent models, with SciBERT often outperforming BERT in tasks on papers outside of its training domain: SciBERT has performed well in shared tasks in scholarly publication processing such as \cite{guangyuan2021} and \cite{10.1007/978-3-031-65794-8_16}, and in studies such as \cite{ming2020} and \cite{poleksic2023-ceurws} has demonstrated superior performance compared to BERT for scientific domain specific tasks.

\subsection{The Current Study}\label{sec:current-study}

In this study, we extend our previous paper \citep{culbertISSI2025} by replicating the previous study on a new and balanced dataset, and compute the DSI of the combined titles and abstracts of papers contained within Clarivate's Web of Science (WoS\footnote{\url{https://clarivate.com/academia-government/scientific-and-academic-research/research-discovery-and-referencing/web-of-science/}}) from a diverse number of fields and over time, to explore whether there exist trends in originality that correlate with field of research, primary subject classification, number of authors, publication date, or citation count. 
We further this by computing DSI scores using multiple embedding models and comparing the correlations between them, as well as examining their sensitivity to two bibliometric variables: author count and publication year.

We aim to introduce the DSI metric to scientometric research as a quantitative indicator of textual originality. We do so by computing DSI scores for combined titles and abstracts of scientific papers from 80 topics across science (based on WoS) and across a large time span from 1994-2025 using two embedding models: BERT, following \cite{Johnson2023} and our previous paper \cite{culbertISSI2025}, and SciBERT, a novel contribution. In line with \cite{holyst024}, we analyse how DSI scores change over time across diverse scientific topics to determine whether abstract originality has generally increased or decreased.

We hypothesize that: a) the DSI of the combined titles and abstracts of scientific papers will correlate with their citation count when other bibliometric variables are controlled for. Furthermore, given prior findings demonstrating the superiority of SciBERT over BERT for scientific texts, b) we hypothesize that DSI scores computed with SciBERT will exhibit a stronger correlation with citation counts than those computed with BERT. Finally, we perform exploratory analyses on the relation of DSI scores with standard bibliometric metrics: number of authors, field of research, and publication year, c) We expect DSI scores to vary across scientific domains and publication years, and we anticipate that the number of authors will further influence these scores.

\section{Methodology}

\subsection{Data}\label{sec:Data}
We obtained the titles, abstracts, and bibliometric information of scientific papers from the WoS as of July 2025, provided by the Competence Network for Bibliometrics \citep{10.1162/QSS.a.20}. From this database we retrieved all subject categories with over 10,000 records with classification \enquote{Article}.

For all scientific papers analysed, the following bibliometric information was extracted from the Competence Network for Bibliometrics’ version of the WoS: \enquote{Primary Subject}, \enquote{Publication Year}, \enquote{Citations after 3 Years}, \enquote{Citations after 5 Years} and \enquote{Total Citations}. We identified the Field of Research (field) for each primary subject by correlating the Leiden University Centre for Science and Technology's (CWTS NOWT-WoS) classification of the WoS\footnote{\url{https://www.cwts.nl/pdf/nowt_classification_sc.pdf}} and Clarivate's Research Areas\footnote{Originally at \url{https://images.webofknowledge.com/images/help/WOS/hp_research_areas_easca.html}, Accessed Jan 2025, no longer available. Now found at \url{https://webofscience.zendesk.com/hc/en-us/articles/38543541713169-Research-Areas}.},  which is visible in \textcolor{blue}{Figure 4}. Notably in the NOWT classification, the subject Multidisciplinary Sciences was classified into its own field, and we follow this convention, although this leads to a comparatively higher variance for this field due to its smaller size.

In our previous paper \citep{culbertISSI2025}, we selected subjects which have at least 1000 abstracts with 199-299 spaces, which we assumed correlates to 200-300 words in each abstract. The restriction on the number of spaces (and thereby words) was to prevent tokenized texts from overrunning the maximum token limit for BERT and SciBERT.

However, we found that the data was biased towards recent years as the random sample was more likely to pick up papers from the more abundant modern years. 

Our sampling strategy for the current study was to choose 20 subjects per field with a minimum of 20 articles per year from 1994 to 2025, each of which having an abstract with 199-299 spaces (again to respect the maximum token limit of BERT and SciBERT).  Due to a lack of data, only 19 categories from the Social Sciences and a single category from Arts and Humanities were selected. Therefore, we excluded the Arts and Humanities field from the analysis. 
This led to a dataset of a total of 51,200 articles composed of 80 categories: 20 from each of \enquote{Life Sciences \& Biomedicine}, \enquote{Physical Sciences}, \enquote{Technology}, 19 from \enquote{Social Sciences} and the field \enquote{Multidisciplinary Sciences} containing a single subject also named \enquote{Multidisciplinary Sciences}.

\begin{table}[htbp] 
    \centering
    \footnotesize
    \setlength{\tabcolsep}{2pt}
    \makebox[\textwidth][c]{
        \begin{tabular}{lr|rrrrr|rrrr|rrrr|rrrr}
            \toprule
            & & \multicolumn{5}{c}{\textbf{Authors}} & \multicolumn{4}{c}{\textbf{3yr Cit.}} & \multicolumn{4}{c}{\textbf{5yr Cit.}} & \multicolumn{4}{c}{\textbf{All Cit.}} \\
            \cmidrule(lr){3-7} \cmidrule(lr){8-11} \cmidrule(lr){12-15} \cmidrule(lr){16-19}
            \textbf{Field} & \textbf{N} & \rotatebox{90}{\textbf{Mean}} & \rotatebox{90}{\textbf{Med.}} & \rotatebox{90}{\textbf{Max.}} & \rotatebox{90}{\textbf{S.D.}} & \rotatebox{90}{\textbf{Zeros}} & \rotatebox{90}{\textbf{Mean}} & \rotatebox{90}{\textbf{Med.}} & \rotatebox{90}{\textbf{S.D.}} & \rotatebox{90}{\textbf{Zeros}} & \rotatebox{90}{\textbf{Mean}} & \rotatebox{90}{\textbf{Med.}} & \rotatebox{90}{\textbf{S.D.}} & \rotatebox{90}{\textbf{Zeros}} & \rotatebox{90}{\textbf{Mean}} & \rotatebox{90}{\textbf{Med.}} & \rotatebox{90}{\textbf{S.D.}} & \rotatebox{90}{\textbf{Zeros}} \\
            \midrule
            Life Sci. \& Biomed. & 12,800 & 5.63 & 5 & 177 & 4.58 & 2 & 5.16 & 3 & 7.73 & 2,398(18.7) & 10.1 & 6 & 15.6 & 1,402(10.9) & 30.3 & 15 & 55.9 & 832(6.5) \\
            Multidisciplinary & 640 & 6.14 & 5 & 100 & 5.18 & 0 & 14.5 & 7 & 22.6 & 71(11.1) & 28.7 & 14 & 45.5 & 55(8.59) & 96.3 & 33 & 195 & 38(5.94) \\
            Physical Sci. & 12,800 & 7.7 & 3 & 2,957 & 85.6 & 1 & 5.8 & 3 & 9.54 & 2,124(16.6) & 10.8 & 6 & 17.5 & 1,353(10.6) & 32.7 & 14 & 93.6 & 868(6.78) \\
            Social Sci. & 12,160 & 2.94 & 2 & 50 & 2.1 & 3 & 3.55 & 2 & 6.11 & 3,184(26.2) & 7.79 & 4 & 13.7 & 1,842(15.2) & 34 & 13 & 111 & 965(7.94) \\
            Technology & 12,800 & 3.79 & 3 & 92 & 2.58 & 8 & 4.91 & 2 & 8.43 & 3,078(24.1) & 9.86 & 5 & 16.8 & 1,917(15) & 35.2 & 13 & 113 & 1,064(8.31) \\
            \midrule
            \textit{All Fields} & 51,200 & 5.05 & 3 & 2,957 & 42.9 & 14 & 4.99 & 3 & 8.51 & 10,855(21.2) & 9.88 & 5 & 16.8 & 6,569(12.8) & 33.8 & 14 & 98.2 & 3,767(7.36) \\
            \bottomrule
        \end{tabular}
    }
    \caption{Author and citation statistics by field of research. N = number of papers; Author columns: mean, median, maximum, standard deviation, Zeros = papers with zero authors; Citation columns: mean, median, standard deviation, and Zeros = number (percentage) of uncited papers for the three citation accumulation periods}
    \label{tab:compact_stats}
\end{table}

The dataset contains 51,200 articles collected over the 32-year period from 1994 to 2025, with 1600 papers collected per year distributed equally over the 80 subject categories. Therefore, 20 papers per subject per year and a total of 640 papers per subject category. Of these scientific papers, 4,876 (9.523\%) did not contain a Digital Object Identifier (DOI). Summaries of the citation data and author count per paper in the dataset can be found in Table \ref{tab:compact_stats}. In Appendix \ref{appx:subjects}, Table \ref{tab:subjects-by-field} lists each subject and their associated field of research.

For modelling citations after 5 years, we dropped the scientific papers published prior to 2020 to prevent bias from a lack of time to accumulate. For the improved model, we further excluded 14 papers with 0 in the authors count. This resulted in a database of 41,600 articles for the simple model and 41,586 articles for the improved model.

\subsection{Measuring Abstract Originality with DSI}\label{sec:DSI-description}
DSI is the arithmetic mean of the pairwise cosine distance of the embeddings (produced by BERT \citep{devlin2019bertpretrainingdeepbidirectional} in hidden layers 6 and 7) of the sentences in a text with each other. The cosine distance is defined as one minus the inner product of the two input vectors. Mathematically, this can be formulated as: for a given text $T$ represented as an ordered list of length $n>2$ containing sentences $s_{i}$, and the embedding vector from the BERT model at layer $k$ defined as $BERT_{k}(s_{i})=\beta_{(s_i,k)}$:
 \begin{equation}
    DSI([s_{1},s_{2},\ldots,s_{n}])=\sum_{k_{1},k_{2}\in\{6,7\}}\sum_{1 \leq i < j \leq n}\frac{1-\frac{\beta_{(s_i,k_{1})}\cdot\beta_{(s_j,k_{2})}}{\|\beta_{(s_i,k_{1})}\|\cdot\|\beta_{(s_j,k_{2})}\|}}{4n} 
 \end{equation}\label{eq:DSI}

To investigate whether SciBERT is a better embedding model for calculating DSI of scientific papers, we computed the DSI using the cased AllenAI's SciBERT model \citep{beltagy-etal-2019-scibert} as embedding model. Hereafter unless otherwise stated, DSI computed using SciBERT as the embedding model is referred to as SciBERT-DSI and DSI computed with BERT as the embedding model is referred to as BERT-DSI, if DSI is stated without reference to model it is referring to both models.

When computing embeddings using a language model, the text must first be tokenised, this is the process that assigns a number to each wordpiece which is then used by the language model to compute the semantic embedding of the text. Following \cite{Johnson2023} we use the Punkt Sentence Tokeniser from the Python Natural Language Tooklkit (nltk) package \citep{loper2002nltknaturallanguagetoolkit}, which has been trained on the corpus of texts being fed in to segment our input texts into sentences. We then used the appropriate model tokeniser to tokenise each sentence; to prevent errors, we enabled truncation of the tokenised text if the resulting list of tokens was greater than the model maximum token input (in case of BERT and SciBERT, this was 512 tokens.)
 
To allow the faster calculation of DSI, we adapted the code provided alongside \cite{Johnson2023} to run on a GPU. The details of this are discussed in Appendix \ref{appx:gpu}.
The code to compute DSI on GPU is linked in Section \ref{sec:availability}.

\subsection{DSI Computation Illustration}

To clarify this computational procedure, we illustrate the result using two examples, and the computation of BERT-DSI with one example. The examples were selected as the articles with the highest and lowest BERT-DSI values within the primary subject exhibiting the greatest range in our dataset: Mycology. The two works quoted are \cite{WOS:000620231200001} and \cite{WOS:000552521900001} respectively.

\begin{quotebox}[h]
    \begin{tcolorbox}[colback=gray!10, colframe=gray!50, title={Title and Abstract of \cite{WOS:000620231200001}}, fontupper=\tiny]
        \textbf{`Multi-aged forest fragments in Atlantic France that are surrounded by meadows retain a richer epiphyte lichen flora.'}\\`This project was focused on identifying the effect of environmental factors on epiphytic lichen species by using a multiscale design applied within multi-aged forest fragments. The field investigations were performed within 20 forest fragments, of which 14 were surrounded by crops and six were surrounded by meadows. Sampling units of 10 by 10 m were selected from the exterior to the interior of each forest fragment following the perimeter line; other sampling units were selected following the same perimeter line to the centre of the forests. The spatial gradient represented by the exterior and interior parts of the forest fragments, surrounding matrix and forest structure (i.e., the presence of larger trees) significantly supported patterns of lichen abundance and diversity. Lichen abundance and diversity were significantly influenced by microhabitat and macrohabitat drivers on the relatively large trees in the forest fragments surrounded by both crops and meadows. Lichen species replacement was significantly described by both larger and thinner trees situated in the interior and at the exterior of the forest fragments surrounded by meadows. The lichen richness was significantly higher on larger trees situated in the interior of the forest fragments surrounded by meadows. The mature structure of forests and the surrounding matrix significantly determined the pattern of epiphytic lichen species. Furthermore, larger and thinner trees harbour very rare lichen species within forest fragments surrounded by both crops and meadows. Forest management practices based on selective cutting on a short rotation cycle did not exert a negative impact on epiphytic lichen.'
    \end{tcolorbox}
    \label{quote:Vicol-low-DSI}
    \caption{Low BERT-DSI Example. Title and abstract from \cite{WOS:000620231200001}, BERT-DSI = 0.563; SciBERT-DSI = 0.630}
\end{quotebox}

\begin{quotebox}[h]
    \begin{tcolorbox}[colback=gray!10, colframe=gray!50, title={Title and Abstract of \cite{WOS:000552521900001}}, fontupper=\tiny]
        `\textbf{Culturable mycobiota from Karst caves in China II, with descriptions of 33 new species.'}\\`Karst caves are characterized by darkness, low temperature, high humidity, and oligotrophic organisms due to its relatively closed and strongly zonal environments. Up to now, 1626 species in 644 genera of fungi have been reported from caves and mines worldwide. In this study, we investigated the culturable mycobiota in karst caves in southwest China. In total, 251 samples from thirteen caves were collected and 2344 fungal strains were isolated using dilution plate method. Preliminary ITS analyses showed that these strains belonged to 610 species in 253 genera. Among these species, 88.0\% belonged to Ascomycota, 8.0\% Basidiomycota, 1.9\% Mortierellomycota, 1.9\% Mucoromycota, and 0.2\% Glomeromycota. The majority of these species have been previously known from other environments, and some of them are known as mycorrhizal or pathogenic fungi. About 52.8\% of these species were discovered for the first time in karst caves. Based on morphological and phylogenetic distinctions, 33 new species were identified and described in this paper. Meanwhile, one new genus ofCordycipitaceae,Gamszarea, and five new combinations are established. This work further demonstrated that Karst caves encompass a high fungal diversity, including a number of previously unknown species. Taxonomic novelties: New genus:GamszareaZ.F. Zhang \& L. Cai; Novel species:Amphichorda cavernicola,Aspergillus limoniformis,Aspergillus phialiformis,Aspergillus phialosimplex,Auxarthron chinense,Auxarthron guangxiense,Auxarthronopsis globiasca,Auxarthronopsis pedicellaris,Auxarthronopsis pulverea,Auxarthronopsis stercicola,Chrysosporium pallidum,Gamszarea humicola,Gamszarea lunata,Gamszarea microspora,Gymnoascus flavus,Jattaea reniformis,Lecanicillium magnisporum,Microascus collaris,Microascus levis,Microascus sparsimycelialis,Microascus superficialis,Microascus trigonus,Nigrospora globosa,Paracremonium apiculatum,Paracremonium ellipsoideum,Paraphaeosphaeria hydei,Pseudoscopulariopsis asperispora,Setophaeosphaeria microspora,Simplicillium album,Simplicillium humicola,Wardomycopsis dolichi,Wardomycopsis ellipsoconidiophora,Wardomycopsis fusca; New combinations:Gamszarea indonesiaca(Kurihara \& Sukarno) Z.F. Zhang \& L. Cai,Gamszarea kalimantanensis(Kurihara \& Sukarno) Z.F. Zhang \& L. Cai,Gamszarea restricta(Hubka, Kubatova, Nonaka, Cmokova \& \& x158;ehulka) Z.F. Zhang \& L. Cai,Gamszarea testudinea(Hubka, Kubatova, Nonaka, Cmokova \& \& x158;ehulka) Z.F. Zhang \& L. Cai,Gamszarea wallacei(H.C. Evans) Z.F. Zhang \& L. Cai.
    \end{tcolorbox}
    \label{quote:zhang-high-DSI}
    \caption{High BERT-DSI Example. Title and abstract from \cite{WOS:000552521900001}, BERT-DSI = 0.700; SciBERT-DSI = 0.705}
\end{quotebox}

For the low BERT-DSI example, the computation begins by splitting the text into sentences using the Punkt Sentence Classifier, which is pre-trained on all texts within the primary subject. This would break the text down into a list of sentences such as:

\begin{quote}
[`Multi-aged forest fragments in Atlantic France that are surrounded by meadows retain a richer epiphyte lichen flora.', `This project was focused on identifying the effect of environmental factors on epiphytic lichen species by using a multiscale design applied within multi-aged forest fragments.', ..., `Forest management practices based on selective cutting on a short rotation cycle did not exert a negative impact on epiphytic lichen.']    
\end{quote}

This list of sentences is referred to in Formula \ref{eq:DSI} as $T$, and each sentence as $s_{i}$ where $i$ is the index of the list.

This list would then be tokenised, which takes each sentence $s_{i}$ and converts it into a list of tokens which correspond to numbers which the embedding model can operate on. In the case of SciBERT and BERT, the tokeniser is a word-part tokeniser which means that it will split words into constituent parts. For example, the first sentence $s_1$ is tokenised as:
\begin{quote}
    ['[CLS]', 'Multi', '\#\#aged', 'forest', 'fragments', 'in', 'Atlantic', 'France', 'that', 'are', 'surrounded', 'by', 'meadows', 'retain', 'a', 'rich', '\#\#er', 'e', '\#\#pi', '\#\#phy', '\#\#te', 'l', '\#\#iche', '\#\#n', 'flora', '[SEP]']
\end{quote}

Which is represented numerically as:
\begin{quote}
[101, 18447, 15841, 3304, 11062, 1107, 3608, 1699, 1115, 1132, 4405, 1118, 25958, 8983, 170, 3987, 1200, 174, 8508, 22192, 1566, 181, 26312, 1179, 16812, 102].
\end{quote}

This is then fed as an input into BERT, and embeddings from the 6th or 7th layer would be extracted. The embeddings in a single layer are 26, 1024-entry long vectors (equivalently, lists) of decimal numbers, and represent the input's semantic position within the representation of language learnt by the BERT model, provided by the training process used to generate the internal weights of the model. In the sixth layer, when computing the above tokenised sentence, the first 1024 long embedding vector looks like this:
\begin{quote}
[ 0.2561,  0.0798,  0.0367,  ...,  0.0201, -0.0087, -0.0106]    
\end{quote}

To compute DSI, all 26 of the embeddings from layer 6 and 26 from layer 7 in the model are extracted for each input sentence and concatenated resulting in a embedding vector $\beta_{s_i,k}$, where $s_i$ is the sentence and $k$ is the layer. Then, for each pair of sentences, the average cosine distance (which is a measure of dissimilarity between two high-dimensional vectors, applicable to these two, now 26,624 dimensional, vectors) is computed.

The cosine distance is computed as one minus the cosine similarity, and the formula for the cosine distance $D(v_1,v_2)$,  between two vectors $v_1$ and $v_2$, with the absolute value of the vector being represented as $\|v\|$ and the dot product as $v_1\cdot v_2$ is as follows:

\begin{equation*}
    D(v_1,v_2) = 1 - \frac{v_1\cdot v_2}{\|v_1\|\|v_2\|}
\end{equation*}

Therefore, the algorithm computes the distance between both layers of the first and second sentences:
\begin{quote}
$D(\beta_{s_1,6},\beta_{s_2,6})$, $D(\beta_{s_1,6},\beta_{s_2,7})$, $D(\beta_{s_1,7},\beta_{s_2,6})$, $D(\beta_{s_1,7},\beta_{s_2,7})$
\end{quote}
These numbers are then computed for the first and third sentences, and so on for all possible pairs of sentences without replacement in the list.
These are then averaged to give a DSI for the entire text $T$, in the case of this text: 0.563.

It should be noted that this number conveys no inherent measure of scientific importance to the text, and is simply a measure of the textual originality.

\subsection{Analysis Approach}\label{sec:analysis-approach}

This study extends our previous paper \citep{culbertISSI2025}, and we aim to reproduce and extend the analyses in the current paper alongside our new contributions.

Our initial work collected and analysed a dataset that was balanced over time and fields, which we describe in Section \ref{sec:Data}. In particular, we sought out anomalous bibliometric metadata from this sample and summarized them in Table \ref{tab:compact_stats}. Finding the anomalies in the dataset acceptable, we then computed DSI as described in Section \ref{sec:DSI-description}.

To determine whether the distribution and properties of DSI changed between the dataset studied in the previous paper \citep{culbertISSI2025} and the current study, we first observed the distribution of DSI for both embedding models across fields of research in Figure \ref{fig:violin-dsi} and Table \ref{tab:dsi-comparison}. This demonstrates that, in essence, DSI behaves robustly with scientific texts. We can also compare the distribution and properties of SciBERT against BERT as an embedding model for DSI using this table and figure.  Following the previous paper \citep{culbertISSI2025}, we will then graph DSI over time (Figure \ref{fig:dsi-over-time}) and boxplots of DSI for each subject (Figure \ref{fig:boxplots}) to observe whether there has been an observable temporal change and whether there are differences per field and subject in abstract DSI, and whether this is mirrored in SciBERT-DSI.

To assess how the strength of the DSI–citation relationship varies with author counts and publication year—that is, the sensitivity of DSI to these variables—we compute and plot Spearman correlations between DSI and citation counts across binned author counts and publication years (\ref{fig:author-sensitivity} and Figures \ref{fig:pubyear-sensitivity}). 

To study the predictive power of DSI on citations, as well as to measure the degree of any potential effect, we constructed a log-linear model of five-year citation counts using DSI and the available bibliometric variables in Section \ref{sec:modelling}. We first recreated the model from \cite{culbertISSI2025} to compare the performance of BERT-DSI on the new dataset.

As approaches to originality and scholarly document processing are beginning to leverage large language models (LLMs) \citep{ORGANISCIAK2023101356, paige-etal-2024-training, taffa_leveraging_2023, orwig2024}, we also ran the above analysis on a third model, Google's Gemini embedding model \citep{geminiteam2025geminifamilyhighlycapable}, specifically \enquote{gemini-embedding-001}\footnote{\url{https://ai.google.dev/gemini-api/docs/embeddings}}. We hypothesized that the generally observed greater capabilities of LLMs such as Gemini may further increase the accuracy of the semantic embedding in the calculation of DSI, and therefore that the measurement of scientific originality may be stronger. 

However, as the Gemini embedding model API only provided a single embedding vector per input text, the formula to calculate DSI had to be adjusted to compensate for this restriction. Furthermore, the dimensionality of the resulting vector was much larger at 3072 dimensions. As this is likely a final layer embedding rather than an hidden layer (as used in the standard calculation of DSI from \cite{Johnson2023}) of the neural network this vector is likely to have different properties. Therefore, as we have no ground-truth to verify the correlation with originality as observed in \cite{Johnson2023}, and as seen in Appendix Figure \ref{fig:gemini-pairwise}, there was a lack of correlation between BERT-DSI, SciBERT-DSI and Gemini-DSI. As such, we only report these analyses in Appendix \ref{appx:gemini}.

\section{Results}
\subsection{BERT- and SciBERT-DSI Distributions}\label{sec:dsi-distributions}
\begin{figure}[htbp]
    \centering
    \includegraphics[width=\linewidth]{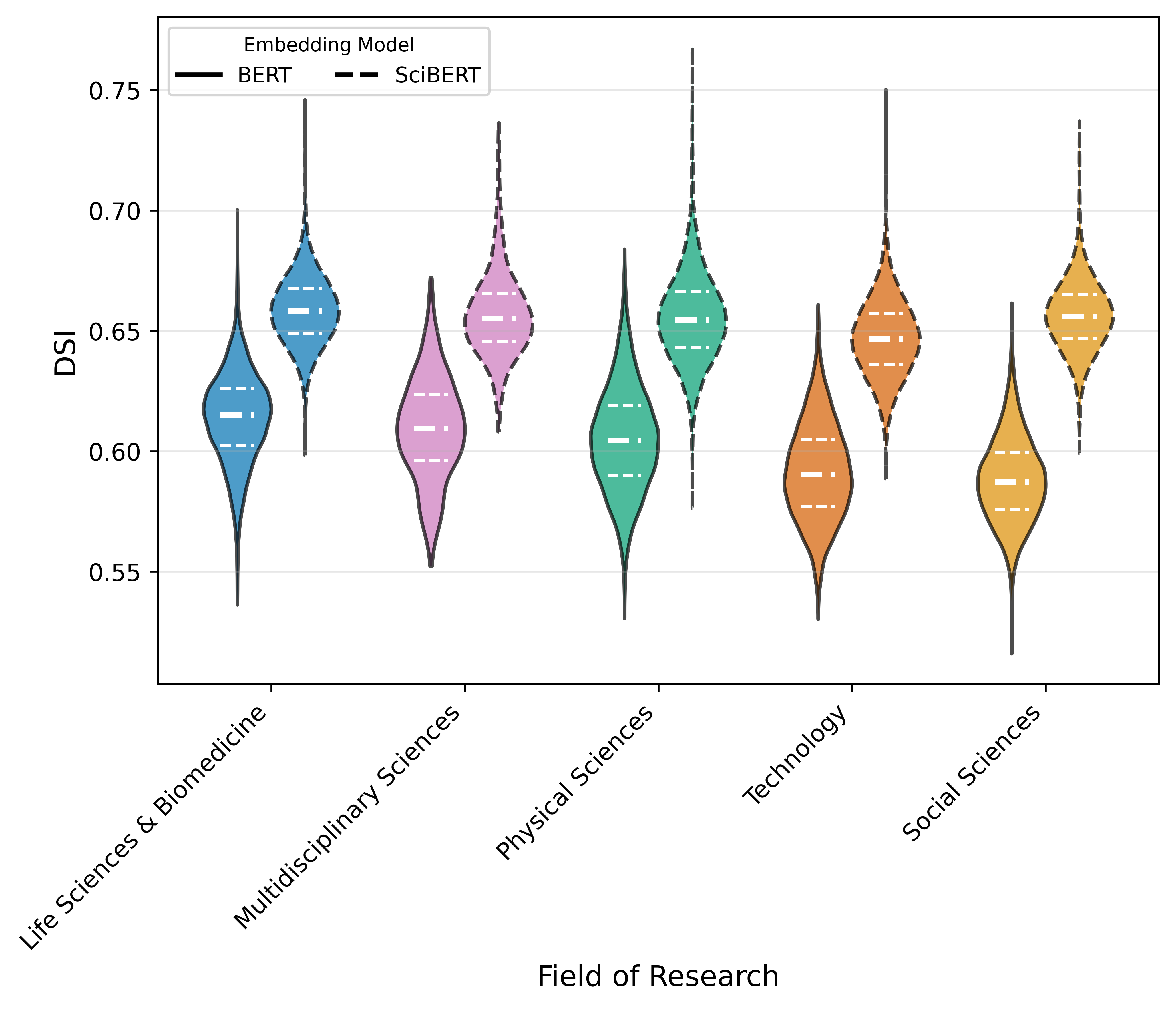}
    \caption{Violin plots of the DSI for each field, ordered by mean BERT-DSI}
    \label{fig:violin-dsi}
\end{figure}

The distribution of DSI by model and domains of science is visualised in Figure \ref{fig:violin-dsi} and tabulated in Table \ref{tab:dsi-comparison}.

\begin{table}[htbp] 
    \centering
    \footnotesize
    \makebox[\textwidth][c]{
        \begin{tabular}{llrrrrrrrrr}
            \toprule
            \textbf{Model} & \textbf{Field of Research}  & \textbf{Min} & \textbf{Q1} & \textbf{Median} & \textbf{Mean} & \textbf{Q3} & \textbf{Max} & \textbf{Range} & \textbf{SD} \\
            \midrule
            \multirow{5}{*}{\textbf{BERT}} 
                & Life Sciences \& Biomedicine & 0.536 & 0.603 & 0.615 & 0.614 & 0.626 & 0.700 & 0.164 & 0.0185 \\
                & Multidisciplinary Sciences & 0.552 & 0.596 & 0.609 & 0.609 & 0.623 & 0.672 & 0.120 & 0.0215 \\
                & Physical Sciences & 0.531 & 0.590 & 0.604 & 0.605 & 0.619 & 0.684 & 0.153 & 0.0218 \\
                & Social Sciences & 0.516 & 0.576 & 0.587 & 0.588 & 0.599 & 0.661 & 0.145 & 0.0177 \\
                & Technology & 0.530 & 0.577 & 0.590 & 0.591 & 0.605 & 0.661 & 0.131 & 0.0201 \\
                \cmidrule(lr){2-10}
                & \textit{All Fields} & 0.516 & 0.583 & 0.599 & 0.600 & 0.616 & 0.700 & 0.184 & 0.0222 \\
            \midrule
            \multirow{5}{*}{\textbf{SciBERT}} 
                & Life Sciences \& Biomedicine & 0.598 & 0.649 & 0.658 & 0.659 & 0.668 & 0.746 & 0.148 & 0.0144 \\
                & Multidisciplinary Sciences & 0.608 & 0.645 & 0.655 & 0.657 & 0.665 & 0.736 & 0.128 & 0.0177 \\
                & Physical Sciences & 0.576 & 0.643 & 0.654 & 0.655 & 0.666 & 0.768 & 0.192 & 0.0181 \\
                & Social Sciences & 0.599 & 0.647 & 0.656 & 0.656 & 0.665 & 0.737 & 0.138 & 0.0141 \\
                & Technology & 0.589 & 0.636 & 0.647 & 0.647 & 0.657 & 0.750 & 0.161 & 0.0161 \\
                \cmidrule(lr){2-10}
                & \textit{All Fields} & 0.576 & 0.644 & 0.654 & 0.654 & 0.664 & 0.768 & 0.192 & 0.0164 \\
            \bottomrule
        \end{tabular}
    }
    \caption{Summary statistics of DSI computed with BERT and SciBERT, broken down by Field of Research}
    \label{tab:dsi-comparison}
\end{table}

We observe that SciBERT-DSI scores have consistently higher mean values than BERT-DSI scores, accompanied by smaller standard deviations but comparable overall ranges. The distribution is more symmetric and concentrated when computed with SciBERT-DSI, with a defined but not normal bell curve and long tails. The distribution of BERT-DSI is less concentrated, with wider and less evenly distributed, non-symmetric tails.

Comparing this to the distribution found in our previous study \citep{culbertISSI2025} we observe no change in the ordering of mean BERT-DSI values, however the BERT-DSI scores are lower. We checked for homogeneity of variances of DSI between fields using Levene's test for both models - both failed with statistics of 162.3 for BERT-DSI and 191.5 for SciBERT-DSI, therefore we did not perform an ANOVA on the differences in DSI between fields which was performed in the previous study.

SciBERT-DSI being universally higher than BERT-DSI shows that the average SciBERT embedding of sentences in academic texts is more widely dispersed in the high dimensional embedding space than the average BERT embedding. This is to be expected as the pretraining of SciBERT has given it greater direct exposure to scientific writing than BERT. The lower standard deviation and associated thinner tails of SciBERT-DSI implies that the model is less sensitive and more regular in its embedding of texts than BERT.

\begin{figure}[htbp]
    \centering
    \includegraphics[width=\linewidth]{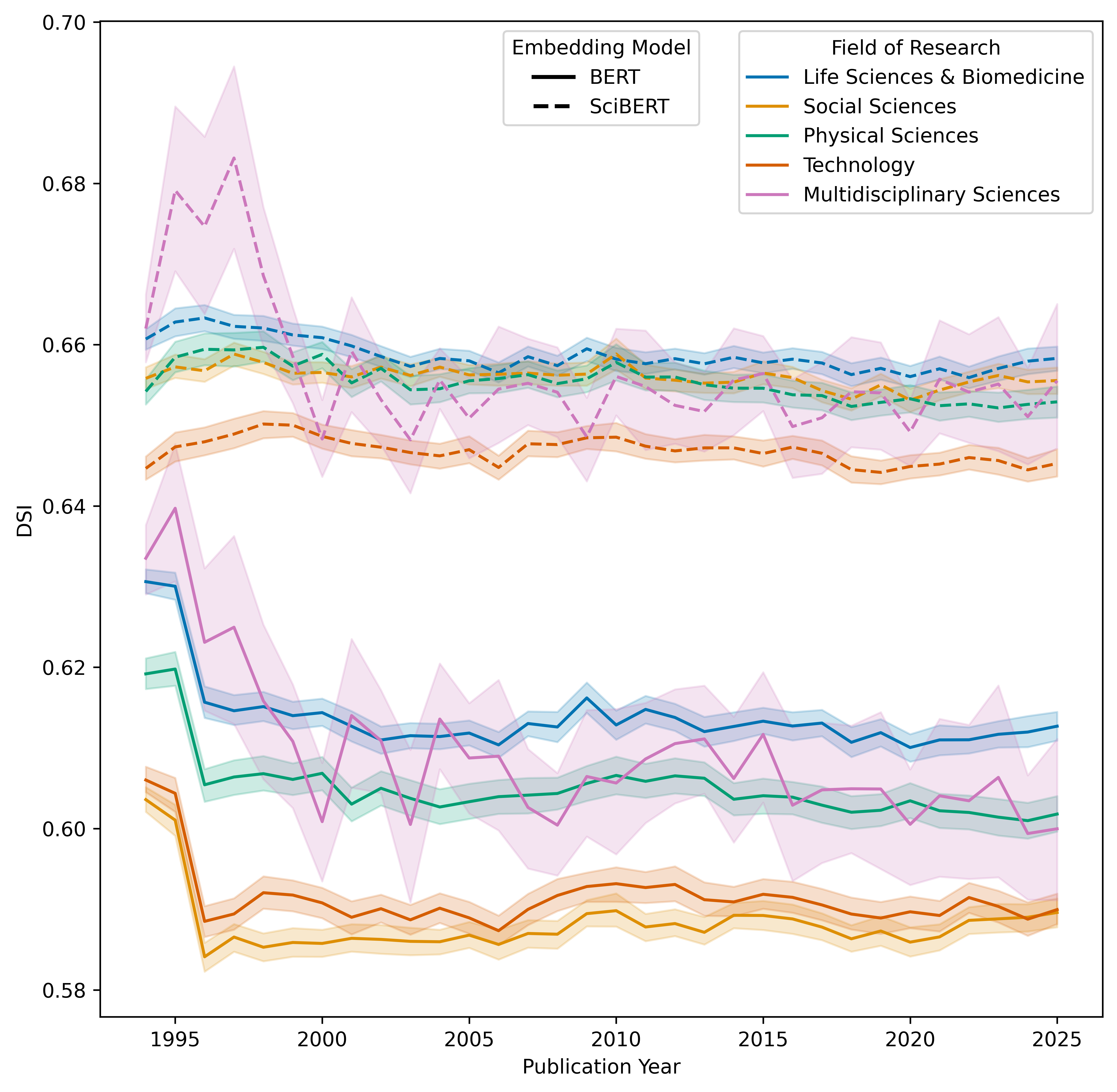}
    \caption{Line plots of DSI by publication year with 95\% confidence interval, by field}
    \label{fig:dsi-over-time}
\end{figure}

Observing the progression of BERT-DSI per scientific domain over time in Figure \ref{fig:dsi-over-time}, we see a higher average DSI in the early 1990s, which falls and remains relatively stable if not trending slightly positive since 1997 for each scientific domain excluding Multidisciplinary Sciences. We have not been able to explain this anomalous drop in BERT-DSI, and we discuss it in detail in Section \ref{sec:discussion-drop}.

For all scientific domains excluding Multidisciplinary Sciences, we do not observe the discussed drop in SciBERT-DSI around 1995-7, visible in Figure \ref{fig:dsi-over-time}. 
This implies that the drop may be an artefact of BERT-DSI, of the textual or collection characteristics within WoS, or of an overall stylistic change in scientific writing at this time. However, evidence to the contrary exists in the dataset used in this paper with the Multidisciplinary Sciences field: which also shows a higher SciBERT-DSI in the 1995-98 range, and then lower afterwards - favouring the textual/collection hypothesis.

We observe a higher SciBERT-DSI than BERT-DSI for nearly all papers analysed, and, when normalised, a linear correlation between the two with Pearson r value $= 0.700$ and p-value $< 0.0001$, which is visualized in in Appendix Figures \ref{fig:gemini-pairwise} and \ref{fig:pairwise-BERT-SciBERT}.

\begin{figure}[htbp]
    \centering
    \includegraphics[width=\linewidth, height=\textheight, keepaspectratio]{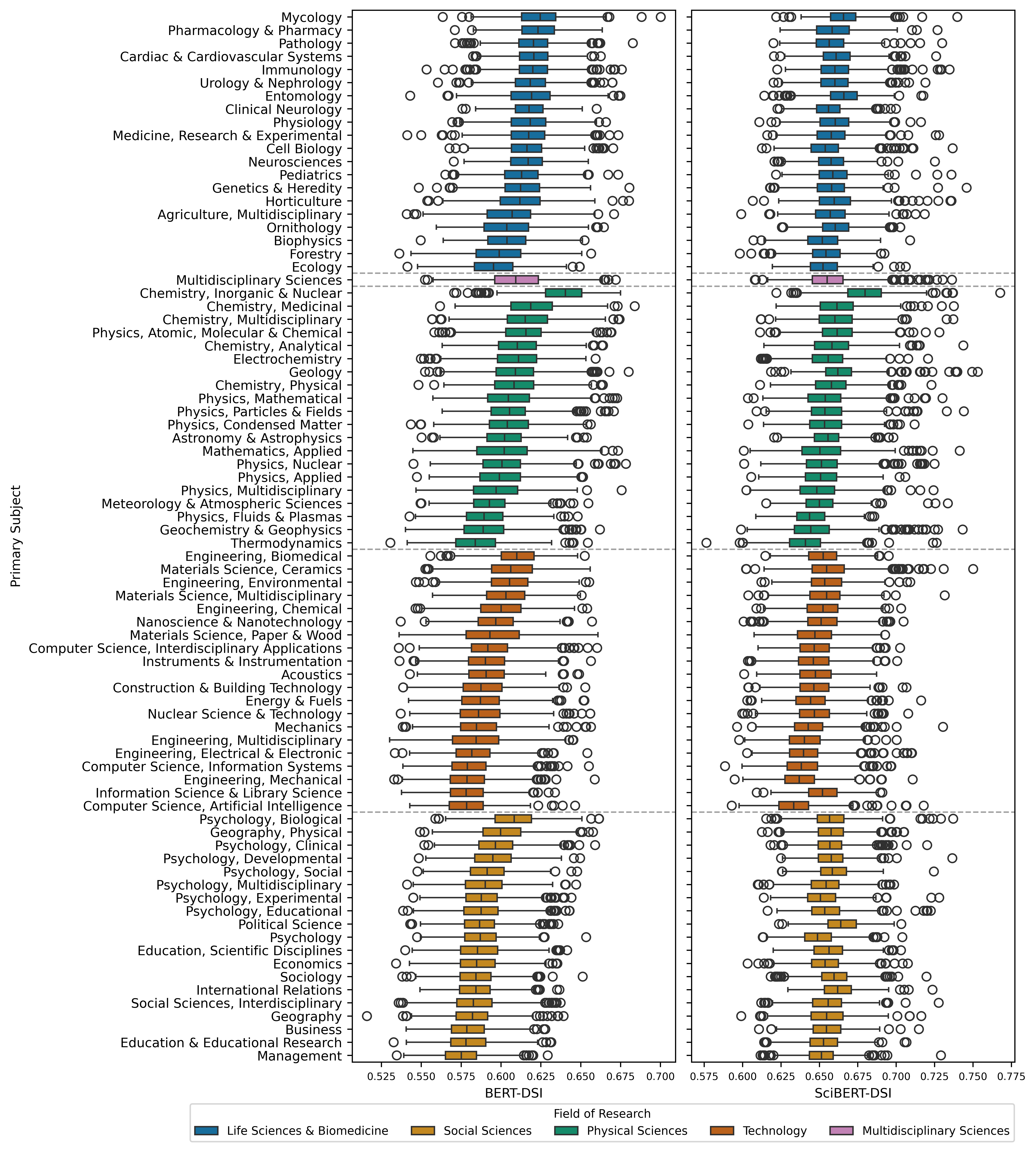}
    \caption{Boxplot of DSI scores per subject, grouped by field of research, ordered by mean BERT-DSI including outliers within each field}
    \label{fig:boxplots}
\end{figure}

In Figure \ref{fig:boxplots}, we break down scientific domains to primary subjects and plot the DSI as a bar chart. We observe broadly similar distributions in DSI across domains: a unimodal bell-curve with thin, long tails and large overlap of the distribution of DSI between subjects and fields. Furthermore, we observe a correlation between mean BERT-DSI and mean SciBERT-DSI broken when grouped by primary subjects (Pearson $r = 0.689$, $p < 0.0001$), where the effect is stronger in Physical Sciences ($r = 0.969$, $p < 0.0001$) and Technology ($r = 0.820$, $p < 0.0001$) and weakest in Life Sciences \& Biomedicine ($r = 0.612$, $p = 0.0041$) and Social Sciences ($r = 0.295$, $p = 0.2195$). In particular, we note that in the domain of Life Sciences \& Biomedicine and the domain of Information Science \& Library Science exceptions to this rule: In Life Sciences \& Biomedicine we observe a visibly weaker correlation between mean BERT-DSI and SciBERT-DSI, and Information Science \& Library Science is anomalously low in the domain of Technology, which otherwise correlates with mean BERT-DSI values very well.

This may imply that BERT-DSI and SciBERT DSI are fundamentally measuring a similar tendency, which supports the motivation of our study of SciBERT as an embedding model. Recalling from \cite{beltagy-etal-2019-scibert}, we note that SciBERT was trained on a random sample of papers: \enquote{18\% from the computer science domain and 82\% from the broad biomedial domain.} The fact that the field which overlaps most with the dataset used in SciBERT's training dataset-Life Sciences \& Biomedicine-does not follow the trend in correlation with BERT-DSI is notable, and the fact that the subject Information Science \& Library Science also behaves anomalously seems to strengthen this observation. However, the primary domains in our dataset seem to more closely align with the computer science domain from the SciBERT training corpus \enquote{Computer Science, Information Science} and \enquote{Computer Science, Interdisciplinary Applications} do not seem to behave anomalously.

\subsection{DSI and Author Count Sensitivity Analysis}\label{sec:author-sensitivity}

Next, we examine how DSI scores relate to author counts. When we studied the dataset, we noticed that some categories have papers with zero author counts, which is obviously incorrect, this is captured in the zeros sub-column in the authors column of Table \ref{tab:compact_stats}. However, given the low absolute number of occurrences (14) we deemed this a negligible database error and continued with this analysis without removing them from the dataset.

\begin{figure}[hbtp]
    \centering
    \includegraphics[width=1\linewidth]{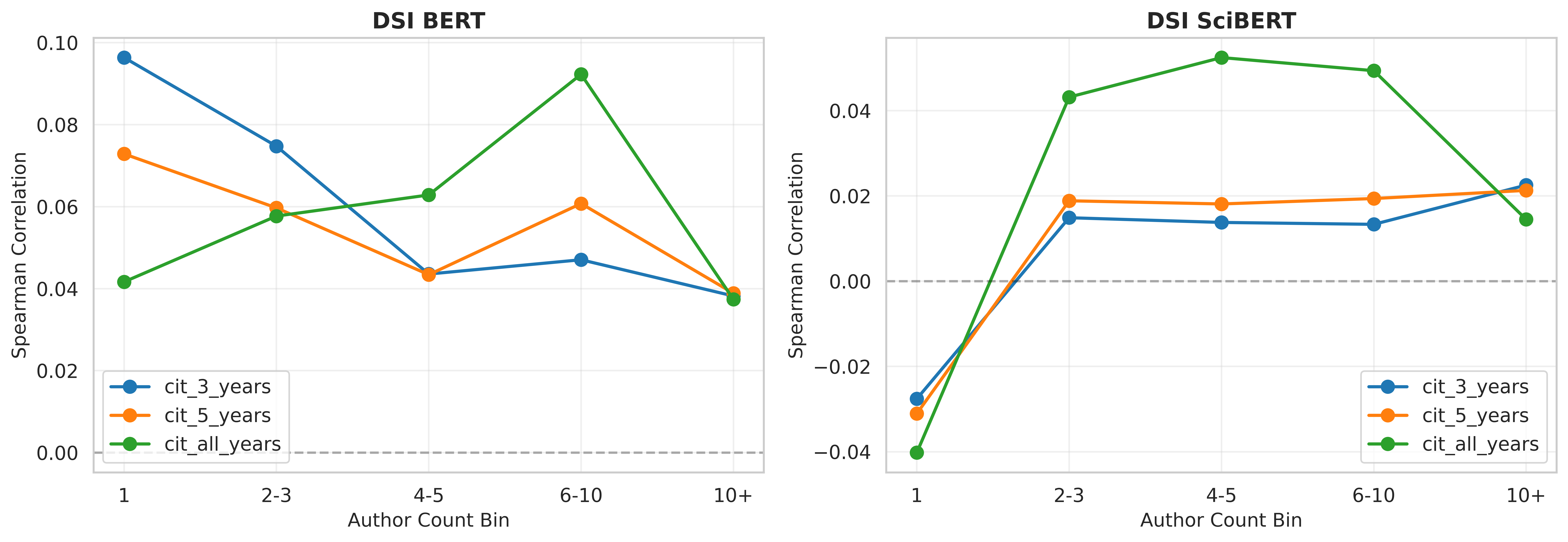}
    \caption{Correlation of DSI with citation counts over binned number of authors for both models}
    \label{fig:author-sensitivity}
\end{figure}

We then examined how DSI correlated with the number of authors of a paper. To do this we binned the author counts (as the range of author group size is very high and contains a few large outliers, particularly in Physical Sciences where group sizes can be in the thousands) we then measured the Spearman correlation between the DSI value and the citation counts in the dataset and plotted the results in Figure \ref{fig:author-sensitivity}.

We observed that BERT-DSI correlates positively with citation count over the author ranges. BERT-DSI shows a higher correlation of DSI to citation score with lower author counts for time limited ranges. However, the opposite is true with medium sized author groups and unrestricted citation counts. 

SciBERT-DSI correlates negatively with citation counts for single author papers, but positively with multi-author papers. SciBERT-DSI also demonstrates little sensitivity to author counts for time limited citation accumulation ranges, but a drop-off is observed in the large author count papers for total citations.

Overall BERT-DSI has a greater absolute value of Spearman correlation to citation counts for each author count bin.

\subsection{DSI and Publishing Year Sensitivity Analysis}\label{sec:pubyear-sensitivity}

In a similar manner to Section \ref{sec:author-sensitivity}, we examine the sensitivity of DSI to scientific papers' publishing year. We correlated the DSI for each model with the citation counts for publications in five 6-year ranges, covering the whole dataset. Please note that the most recent range is plotted for citations after 3 and 5 years, in which there will be papers too recent to have filled the accumulation period. 

\begin{figure}
    \centering
    \includegraphics[width=1\linewidth]{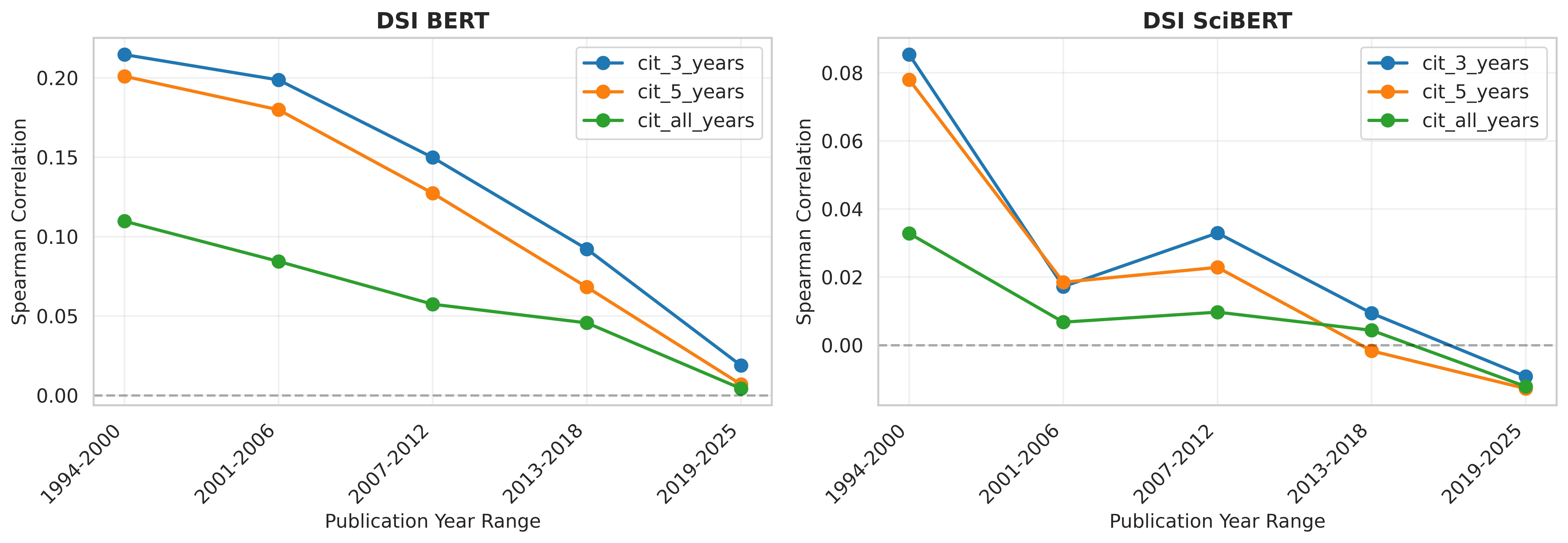}
    \caption{Correlation of DSI with citation counts over publication year ranges for both models}
    \label{fig:pubyear-sensitivity}
\end{figure}

The results are plotted in Figure \ref{fig:pubyear-sensitivity}, in which we see that BERT-DSI has a positive correlation with the citation counts over all years. However, the strength of the predictive power of DSI on citation count is monotonically decreasing across all citation counts. Furthermore, the correlation is clearly stronger for shorter citation-accumulation periods. SciBERT-DSI exhibits a less regular decline in correlation strength, and no clear pattern emerges linking model performance to the length of the citation-accumulation period. In the most recent time period SciBERT-DSI correlates slightly negatively with citation counts across all accumulation periods, and in the second most recent period only citations after 5 years correlate negatively, however the strength of the correlation is very weak.

\subsection{Modelling Scientific Papers' Citation Count Based on DSI}\label{sec:modelling}

Following our previous paper \citep{culbertISSI2025}, we wished to model whether DSI has an effect on scientific impact, as such we modelled citation count using a generalised linear model of DSI and field as a categorical variable. We mitigated the bias due to accrual of citations by older papers by correlating the number of citations after 5 years. Thus, for this model we considered only papers published before the end of 2019, to allow for a fair accrual of 5 years of citations before the 2025 sample date. This restriction left us with a dataset of 41,600 records.

\begin{figure}
    \centering
    \includegraphics[width=1\linewidth]{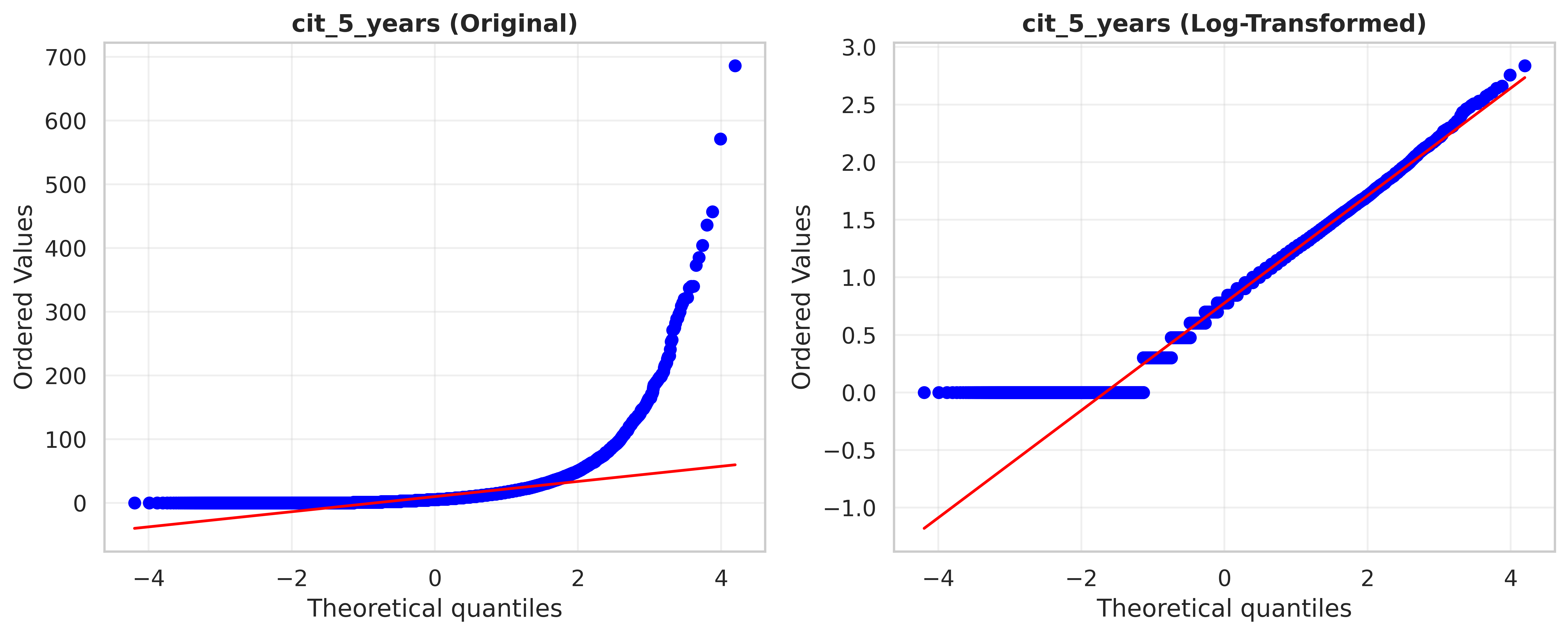}
    \caption{Quantile-Quantile (QQ) plot of untransformed citation data against log transformed citation data}
    \label{fig:qq-log-vs-untransformed}
\end{figure}

As some domains had a large range in citation count after 5 years, and to better model the large differences in average citation count after 5 years by subject, we took the base 10 logarithm of the citation count after 5 years (after adding 1 to all citation counts to prevent the logarithm function mapping to negative infinity for papers with no citations). In Figure \ref{fig:qq-log-vs-untransformed}, Quantile-Quantile plots of the multilinear model predicting citations after 5 years and the log transformed citations after 5 years demonstrate the better fit of the model: $log_{10}(cit_{5 years})+1) \sim DSI+C(Field)$ using log transformed citation counts.

\begin{figure}
    \centering
    \includegraphics[width=\linewidth]{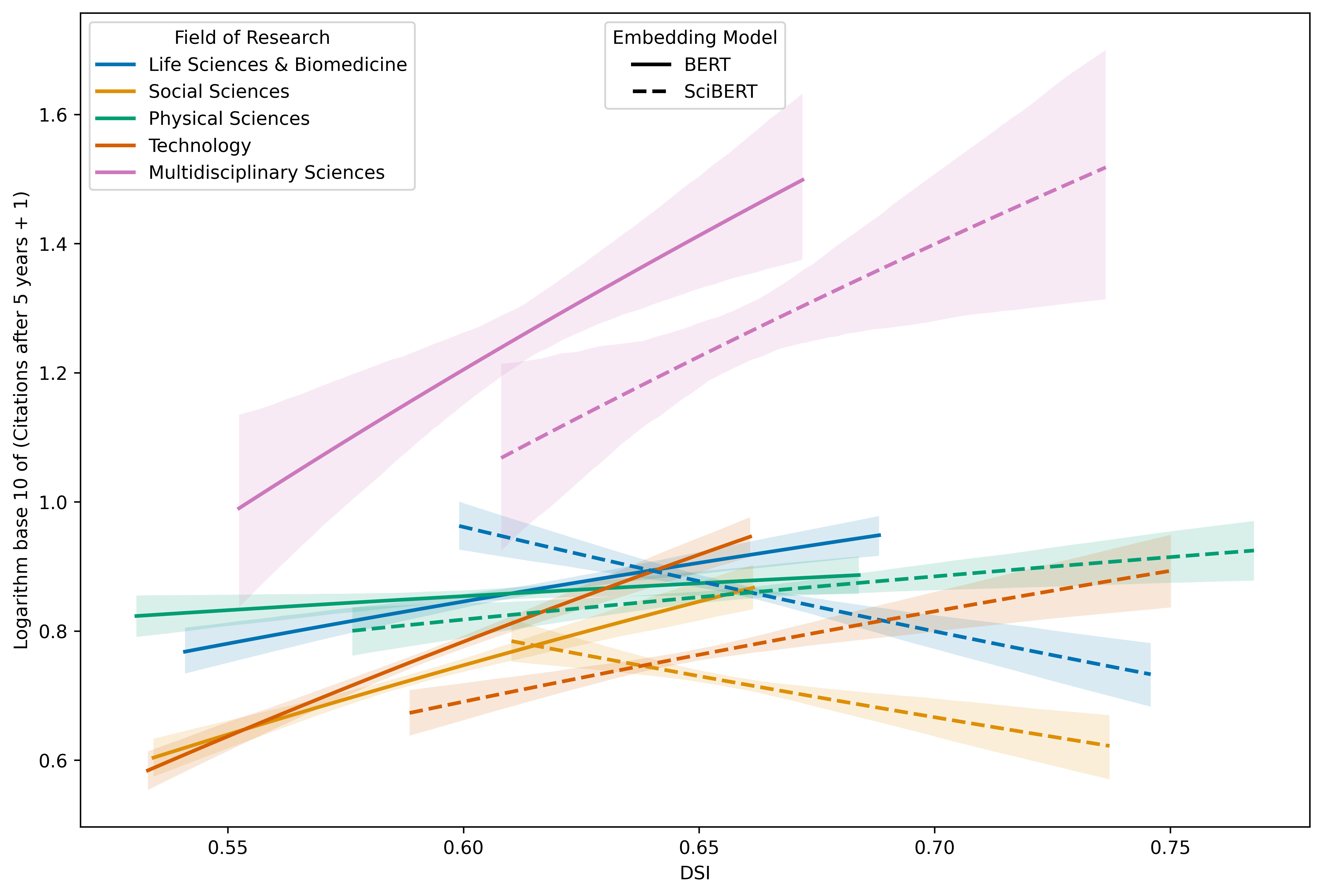}
    \caption{Ordinary Least Squares Regression for the base 10 of the number of citations after 5 years (plus one) predicted by DSI and field for both embedding models, plotted with 95\% confidence interval}
    \label{fig:least-squares}
\end{figure}

\begin{table}[htbp]
    \centering
    \makebox[\textwidth][c]{
        \begin{tabular}{lllrccccccc}
            \toprule
            & & & & \multicolumn{3}{c}{\textbf{Overall Model}} & \multicolumn{2}{c}{\textbf{DSI Predictor}} & \multicolumn{2}{c}{\textbf{Diagnostics}} \\
            \cmidrule(lr){5-7} \cmidrule(lr){8-9} \cmidrule(lr){10-11}
            \textbf{Model} & \textbf{DV} & \textbf{Controls} & $N$ & R$^2$ & $F$ & $p$ & $\beta$ & $p$ & MSE & JB \\
            \midrule
            BERT    & log(cit. 5 years) & Field                     & 41,600 & 0.033 & 287.1 & $<$.001 & 1.594 & $<$.001 & 0.206 & 213.5 \\
            SciBERT & log(cit. 5 years) & Field                     & 41,600 & 0.029 & 246.6 & $<$.001 & 0.056 & .693    & 0.207 & 223.4 \\
            BERT    & log(cit. 5 years) & Field, Year, log(Authors) & 41,586 & 0.103 & 664.9 & $<$.001 & 0.026 & $<$.001 & 0.191 & 131.1 \\
            SciBERT & log(cit. 5 years) & Field, Year, log(Authors) & 41,586 & 0.101 & 646.7 & $<$.001 & 0.002 & .390    & 0.192 & 135.7 \\
            \bottomrule
        \end{tabular}
    }
    \caption{OLS regression results predicting 5-year citations (log-transformed). DV = dependent variable; MSE = Mean Squared Error; JB = Jarque-Bera statistic. Field = Field of Research; Year = publication year (standardised); log(Authors) = number of authors (log-transformed, standardised). Robust standard errors (HC3) used for models with full controls}
    \label{tab:ols_results}
\end{table}
We performed a statistical analysis of the statistical model: $log_{10}(cit_{5 years})+1) \sim DSI+C(Field)$ for DSI computed with both embedding models, both overall models were found to be statistically significant by two-tailed hypothesis test at 99\% confidence. The details of this log-linear model can be found rows 1 and 2 of Table \ref{tab:ols_results} and a regression plot for the model can be found in Figure \ref{fig:least-squares}. The large Jarque-Bera statistics \citep{18bce95a-04fb-37a4-8c3e-29e9bbcae8ac} imply that there is a non-normality to the residuals (errors in prediction) of the model, which can be observed in the Q-Q plot in the lower left as the large number of non-cited papers skews the data above the line of agreement.

To improve this model, as highlighted in the future work section of our previous paper \citep{culbertISSI2025}, we incorporated the other two bibliometric data available: publishing year and author count. 

We then standardised the non-categorical variables-observing the variables we discovered the high degree of skew present in author count. This was due in part to the large outliers present in Physical Sciences visible in the maximum column and mean $>$ median skew present in the Authors column of Table \ref{tab:compact_stats}. We rectified this with a log transformation after filtering out the 14 occurrences of zero author count prior to the standardisation.
Our final log-linear model: $log_{10}(cit_{5 years}+1) \sim DSI + C(Field) + Pubyear + log_{10}(Author Count)$, was statistically significant at 99\% confidence by two-tailed hypothesis test for both embedding models. The details of the log-linear modelling can be found in rows 3 and 4 of Table \ref{tab:ols_results}. This linear model implies that the final BERT model is able to explain 10.3\% of the variation in the logarithm of citations after 5 years, and the final SciBERT model is able to explain 10.1\% of the variation. However the p-value for the gradient of SciBERT-DSI again fails to be statistically significant at 99\% or 95\%, giving more evidence to the hypothesis that SciBERT is not able to predict the logarithm of citations after 5 years.

Analysing the final models we observe that the BERT-DSI coefficient has a value of 0.0259 and 99\% confidence interval of [0.0194, 0.0323]. The SciBERT-DSI has a value of 0.023 and 99\% confidence interval of [-0.0039, 0.0077]. This implies that under this model, a standard deviation increase of BERT-DSI (+0.0222) corresponds with an increase in actual citations after 5 years of +6.1\%, and a standard deviation increase of SciBERT-DSI (+0.0164) would correspond with a +0.4\% increase in actual citations after 5 years.

\section{Discussion}

In Section \ref{sec:analysis-approach} we gave an outline for steps we would take to measure DSI and introduce it as a novel candidate measure of scientific originality.

We began by replicating the analyses within \cite{culbertISSI2025} which we discuss in Section \ref{sec:prevpaper}, but on resampled, smaller dataset, where the number of articles studied in \cite{culbertISSI2025} was 99,577, and we studied 51,200 articles in this paper. 

The reason for resampling was an imbalance across publication year which we believed to be the cause of an unexplained drop in BERT-DSI from the early-mid 1990s in all fields around 1995-7. This turned out to not be the cause and it remained, we discuss this in Section \ref{sec:discussion-drop}.

We first produced an overview of the dataset and the bibliometric variables contained within in Section \ref{sec:Data}. In this section we justified the sampling strategy and discussed obvious errors in the database (such as articles with zero as their author count), and the adjustments made to correct for them.

Our next step was computing DSI with both embedding models, firstly we analysed the differences between BERT-DSI and the computation of BERT-DSI in the previous paper \citep{culbertISSI2025}, which we will cover in detail in Section \ref{sec:prevpaper}. Secondly we looked at SciBERT-DSI and compared this to BERT-DSI, which we discuss in more depth in Section \ref{sec:scibert-analysis}. 

Following this we explored integrating a LLM model as a embedding model for DSI, unfortunately this effort was relatively unsuccessful as the LLM model of choice, Google's Gemini embedding model created DSI scores which were uncorrelated with both BERT-DSI and SciBERT-DSI. We have summarised our findings of this in Appendix \ref{appx:gemini}.

Subsequently we investigated the sensitivity of DSI to publication year and author count, and compared results between embedding models. We found notable trends in BERT-DSI but not SciBERT-DSI, leading us to hypothesise as to the reasons for this and also the implication that our findings brought up in Section \ref{sec:discussion-time-and-author-count}.

We then looked at modelling citations after 5 years to replicate the analysis in our previous paper \cite{culbertISSI2025}. We first used the same model as in our previous paper, where a log-linear model of citations after 5 years was predicted by DSI with controls for field of research. This model had a significant drop in R$^2$ value which we could not explain, however (as outlined in the future work section of \cite{culbertISSI2025}) we planned to model DSI while controlling for more bibliometric variables. When we produced this improved model-with more controls and better modelling-we saw a step up in R$^2$ to approximately the same value as in our previous paper. We discuss this in further detail in Section \ref{sec:disc-modelling}.


\subsection{BERT-DSI Summary} \label{sec:prevpaper}

In \cite{culbertISSI2025}, we previously observed a much higher mean BERT-DSI, this may have been caused by our resampling, as we began sampling in 1994 due to collection rule restraints (the subjects containing at least 20 papers per year from 1980 onwards are dominated by life sciences, from 1994 onwards there are nearly 20 per field which allows us study field of research in this paper). This would cause the observed period of time of higher DSI to not effect our DSI values here. We also observe a drop in relative positioning for Technology, visible in Figure \ref{fig:violin-dsi}, which is now approximately equal in mean DSI to Social Sciences. We continue to observe a positive correlation between DSI and the logarithm of citation count after 5 years, as seen in Figure \ref{fig:least-squares}, although the correlation is weaker.

In our log-linear model $log_{10}(cit_{5 years})+1) \sim DSI+C(Field)$, we found that there was a significant drop in adjusted R$^2$ (from 0.13 to 0.03) and notable increase in Jarque-Bera (from 12.918 in the model from \cite{culbertISSI2025} to 213.5) with the same, simple, model. This implies that there remains a non-normality to the residuals which is visible in Figure \ref{fig:qq-log-vs-untransformed}, where the low ordered values are much higher than the theoretical quantiles. While we observe a decrease in the influence of the DSI value in our final model compared to the simpler model represented by the drop in beta value in Table \ref{tab:ols_results}, this is expected, as correlations with author count and publishing year. Modelling this as a hurdle model (a combination of two models: one to predict whether a paper is cited and one to predict the number of citations if a paper is cited) did not improve on this model. 

We found that standardization of DSI improved the Jarque-Bera and R$^2$ of the log-linear model, and incorporation of standardized publishing year and standardized logarithmically transformed author counts brought the model to an adjusted R$^2$ of 0.10 and Jarque-Bera of 137.6. The remaining large Jarque-Bera implies there is still significant non-normality of the residuals in the dataset. However, the simplicity of the model and fuzzy nature of citation prediction may mitigate this.



\subsection{SciBERT-DSI Summary}\label{sec:scibert-analysis}

The distribution of SciBERT-DSI is more concentrated around the mean (the lower standard deviation for each field and overall is visible in Table \ref{tab:dsi-comparison}) and slightly more symmetrically distributed than BERT-DSI. Figure \ref{fig:boxplots} demonstrates how the loose correlation between BERT-DSI and SciBERT-DSI (further described in Appendix Figure \ref{fig:pairwise-BERT-SciBERT}) extends to a breakdown by subject. We observed that the standard deviation and range of each subject is broadly smaller with SciBERT-DSI in a subject granularity. This correlation seems strongest in Technology and Physical Sciences and weaker in Life Sciences \& Biomedicine and Social Sciences.

When observing the Spearman correlations of SciBERT-DSI and citation counts in Figures \ref{fig:author-sensitivity} and \ref{fig:pubyear-sensitivity}, the general trend of decreasing sensitivity to author count and publishing year observed in the BERT-DSI is not reflected. Further, the correlations are less in magnitude for all bins and ranges. Specifically with number of authors in a paper: we observe a negative correlation between all citation counts and DSI for single author papers, but a positive and fairly stable correlation in all multiple author bins. For publishing year ranges we see a similar descending trend in strength of correlation with SciBERT-DSI to that of BERT-DSI which is weaker in correlation strength. The decreasing trend is less stable with the 2001-2006 year range correlating less strongly than the 2007-2012 range, and enters a negative correlation in the most recent years.

When modelling the logarithm of citation count after 5 years using SciBERT in both the simpler and improved model described in Section \ref{sec:modelling}, we note that the gradient of SciBERT-DSI is not statistically significant. This means that from this dataset, we cannot prove that SciBERT-DSI is a predictor of citations after 5 years. However, when observing the 99\% confidence interval for the effect of SciBERT-DSI in the improved model we see it contains 0, and the fitted effect at 0.015 is much lower than the fitted effect of BERT-DSI at 0.0264 (with 99\% confidence interval [0.0199, 0.0328]). This implies that over this dataset this modelling detects a positive if slight effect of BERT-DSI on log-citations at 99\% confidence but not with SciBERT-DSI. These findings should be taken with the caveat that both SciBERT-DSI and BERT-DSI in this paper have a large Jarque-Bera statistic, which implies that the log-linear model may not be appropriate for predictive inferences of DSI for citation count on this dataset.

These conclusions lead us to hypothesize that the exposure of SciBERT to only these domains' texts within pretraining causes a greater exposure to the subject terms and concepts used within the field. This exposure may train SciBERT to better embed the concepts contained within the scientific data rather than the linguistic and syntactical structures learnt by BERT. This is because BERT uses training data comprising narrative texts from BookCorpus and explanatory texts from Wikipedia rather than scientific texts. This would describe why the SciBERT-DSI and BERT-DSI correlation is so low in Life Sciences \& Biomedicine, however this is not observed in the two Computer Science domains in Technology.

\subsection{Correlation of DSI with Citations over Time and Author Count}\label{sec:discussion-time-and-author-count}
We can infer from Figures \ref{fig:author-sensitivity} and \ref{fig:pubyear-sensitivity} that BERT-DSI and SciBERT-DSI in general are positive predictors of citations by themselves - a purely textual measure of likelihood to be cited, this is with the exception of SciBERT-DSI in single author papers. 
The fact that SciBERT-DSI is correlating less strongly than BERT-DSI in both figures implies that it is a less powerful predictor of citation impact in general, which corroborates the findings of Section \ref{sec:modelling}.

As we observed in Section \ref{sec:pubyear-sensitivity}, the Spearman correlation decreases for more recent papers, and particularly for papers with insufficient citation accrual time in 2020-2025. As these papers are included in our analysis, we note that the effect may in fact be stronger than reported in this study. Furthermore, we see that the BERT-DSI correlation with early citations, after 3 and 5 years, is less powerful with larger author groups. This may signify that trends observed in scientometric research for citation counts to be in general higher for larger author groups \citep{10.1162/qss_a_00003,Adams2005-xp,Tahamtan2016-mr}, are due to factors other than the originality of research as measured by BERT-DSI. 
Equivalently, early citations (known to be a predictor of total citation count \citep{Adams2005-xp}, \cite{Zhang2024-uc}) are predicted more strongly for papers authored by smaller author groups. This indicates that scientific originality is a stronger determining factor for paper impact for smaller author groups than for larger groups.

Across all time ranges except the most recent, BERT-DSI shows a stronger correlation with citation count than SciBERT-DSI. However, because the most recent range includes papers that have not yet had sufficient time to complete their citation-accumulation period, this exception should be interpreted with caution. As above, we note that the degree of Spearman correlation is greater with BERT than SciBERT, further strengthening the case that BERT is the better embedding model for prediction of citations.

The Spearman correlation of citations and DSI is greater in both maximum and minimum magnitude for both models than for author group size. This implies a greater degree of change in the importance of creativity for both a paper's early and total citation counts. As such, the general decreasing trend describes a scientific landscape where one or any of the following trends may occur: One possibility is that the degree of scientific originality is less strongly correlated with citations implying that originality is less important to gaining citations. A second possibility is that the degree of originality in science is less correlated with citations implying more specificity of research and a decrease in the importance of interdisciplinarity and incorporation of concepts from varied fields in research. Finally, other effects such as the availability of academic search engines, databases and scientometric ranking algorithms are driving changes to the textual content of titles and abstracts



\subsection{Modelling Citations after 5 Years with DSI}\label{sec:disc-modelling}
Predicting citations is a known hard challenge in scientometrics. Importantly, such attempts utilise bibliographic information such as journal prestige, journal impact factor and journal language, author H-index and author productivity \citep{BAI2019407, ROBSON201694, 10.1007/s11192-014-1279-6}. Recent work in this area includes analysing citation network structure \citep{ZHAO2022101235}, early citations and open access status \citep{ABRAMO2025101725} or preprint availability \citep{10.1162/qss_a_00043}. In our study, we do not primarily aim to construct an effective citation prediction model but rather study the effect of DSI on scientometric data such as citations, as such we did not explore DSI in context of the advanced models available.

The BERT-DSI model demonstrates that the prediction of citations using DSI and field is able to explain around 3.3\% of the variability in the logarithm of early citation counts. Comparing this to our previous paper \citep{culbertISSI2025} where the identical model is fitted to the previous dataset, we see a significant drop in the predictive power of the model. This may be due to a better representation of older papers, which would imply that there is a shift in BERT-DSI's importance in early citation accrual. This is in fact observed in Figure \ref{fig:pubyear-sensitivity}, and would imply that the model may be stronger on a dataset limited to more recent papers, such as those published 2007-2020 or 2013-2020, as the trend in correlation is changing over time. 

Alternatively, it may be due to the change in distribution of primary subjects, as the resampling of data for our study selected different primary subjects to study (due to the selection criteria) as in our previous paper \citep{culbertISSI2025}. This would imply that the effect of DSI and originality on citations differs significantly by primary subjects within fields, which concurs with the large intra-field variance in DSI by primary subject as observed in Figure \ref{fig:boxplots}. 

The SciBERT model is less powerful than the BERT model in terms of r value, and Figure \ref{fig:least-squares} shows negative effects of SciBERT-DSI on early citation accrual. Notably, this trend is present in Life Sciences \& Biomedicine (as well as Social Sciences) which was unexpected as SciBERT was trained on this field and so one may expect it to be able to perform most accurately in this field. 
However, since the gradient of SciBERT fails to be statistically significant for a 99\% or even 95\% two-tailed hypothesis test gives strong evidence against the hypothesis of SciBERT measuring citation counts. This correlates with the lower Spearman correlations observed in Sections \ref{sec:author-sensitivity} and \ref{sec:pubyear-sensitivity}, the lower standard deviation of DSI as observed in Table \ref{tab:dsi-comparison} and the lack of correlation between BERT-DSI and SciBERT-DSI in Life Sciences \& Biomedicine in \ref{fig:boxplots}. 

Our final models demonstrate a stronger predictive capability than the simpler models, this is partially due to the incorporation of more variables. Comparing between SciBERT and BERT we see the 99\% confidence interval for the gradient of the SciBERT-DSI contains 0 and the 99\% confidence interval for BERT-DSI does not. This highlights that BERT-DSI is a much stronger predictor for citation count. 

The large Jarque-Bera statistics for both models describe non-normal residuals, which means assumptions for linear modelling are not satisfied. This implies that the effect of DSI on citations is non-linear, and that some form of non-linear model may allow for a yet stronger predictive model.

However, this is in contrast to the findings of the previous paper \citep{culbertISSI2025}, the linear model of which did not have significantly non-normal residuals. This invites the same hypotheses and speculation as above as to the reasons for the change in modelling properties. Yet, we can infer from this model that due to the non-negativity of the confidence interval around BERT-DSI that it is a positive predictor of the logarithm of citations after 5 years. Importantly, the effect is statistically significant, which invited further research into DSI as a metric for use in scientometric and bibliometric analyses and its value as a novel purely textual metric for scientific originality.

In summary, we were able to achieve our goal in producing a statistical model which predicts citations after 5 years which measured a statistically significant non-zero effect of BERT-DSI on citation count. This demonstrates that given the limitations of our dataset, model and exploration detailed above and in Section \ref{sec:limitations}, we identified a novel predictor of citations, which we use as a indicator of scientific originality.

\subsection{Observed Drop in BERT-DSI in the Early- to Mid- 1990s}\label{sec:discussion-drop}

After investigating the change in DSI over time in Section \ref{sec:dsi-distributions}, we speculate on the drop in BERT-DSI in the early-mid 1990s. Our investigaton led us to the hypothesis that this is a dataset artefact, as in 1997 the Web of Science was launched - combining indexes (SCI, SSCI and AHCI) formerly compiled by the Institute for Scientific Information, and later Thomson Corporation, now Thompson Reuters. At this time, the data pipeline or data sources may have increased or changed. However, the lack of similar drop in SciBERT-DSI over this time period weakens this hypothesis, and suggests that this may be due to a sensitivity of the BERT model (although SciBERT-DSI Multidisciplinary does show a similar pattern). Furthermore, in our previous paper \citep{culbertISSI2025} it was hypothesized that the higher DSI in the 1980s and early 1990s was due to an undersampling of the dataset used in the paper. We can refute this as the anomaly has remained in the current dataset, although only the end of the anomaly is visible due to the year constraint of 1994 onwards. 


Overall, we do not observe a significant change in DSI with either model since approximately 1997, which implies that the measure is stable across scientific texts and that scientific writing has not been significantly changing in overall textual originality per paper in this time. This is in broad agreement with our previous study \citep{culbertISSI2025}, which reported a very weak positive trend over this time period. However, the change in finding between these studies may be due to the balanced resampling work over time correcting the bias towards more recent papers in the previous study.

\subsection{Limitations}\label{sec:limitations}
As we were aiming to quantify the usefulness of DSI as a measure of scientific originality, we limited ourselves, in modelling citation count, to predicting an assumed indicator of scientific originality. 
However, without an available ground truth--i.e., human ranked originality scores for scientific papers, like done in creativity research \citep{Johnson2023}--it appears to be the best approach.

While we could not control for the English proficiency of the authors of the papers, BERT-DSI was found to generalize across varying cultural and language backgrounds in study 6 of \cite{Johnson2023}. This limitation may influence SciBERT more strongly, although we were unable to quantify this effect in the present paper.

We acknowledge the assumption that DSI generalizes to longer texts is untested - in study 5 of \cite{Johnson2023} DSI was found to stablise after 30-50 words and this was tested up to 200 words, however our dataset contains 200-300 words and therefore lies outside tested parameters.

A contributing factor to the noise in computing DSI may be database and formatting related - as seen in Quote \ref{quote:zhang-high-DSI} the Web of Science has abstracts with uncontrolled UTF-8 characters such as \enquote{\& x158}, improper formatting: \enquote{Novel species:Amphichorda cavernicola,Aspergillus}, and citations and attributions \enquote{Gamszarea indonesiaca(Kurihara \& Sukarno) Z.F. Zhang}. 
While processing such a large dataset these are likely to occur, cleaning such a large dataset was out of scope for this paper. 
Another contributing factor may be the inherent issue that scientific texts use and develop novel phrasing and specialist terminology that is most likely not contained, or rarely contained, in the texts used to train the model, as such the embeddings of these words may be more inaccurate than if the tokeniser and model were trained on scientific texts, or fine-tuned with the new vocabulary.

A limitation to the computation of DSI is also the texts input: scientific originality may not be fully exemplified in the abstract and title alone, and the ideal solution would be the computation of the full text of the paper. However, due to copyright and data access restrictions, gathering this data is difficult. If the data was available for future investigation, we note that with BERT-based models a further restriction in the maximum of input tokens would be a limitation, one that was already run into in a number of combined abstracts and titles considered in this paper. 

As detailed in Appendix \ref{appx:gemini}, in exploring whether DSI computation would be feasible using a LLM, we had to adjust the formula for computation of DSI. 
This was because the APIs for Gemini embeddings did not allow for extraction of hidden layer embeddings of sentences as is possible through BERT and SciBERT. This, and subsequent exploration leading to a lack of correlation with BERT and SciBERT led us to exclude this work from our study.

Our models predicting citations also utilise relatively basic variables and models compared to the complexity of other approaches detailed in Section \ref{sec:creativity}. This may explain why our correlations between DSI and citations are relatively low. This may be due to a number of reasons, such as the dataset available in the WoS is noisy and contains a modality of text not considered by \cite{Johnson2023}. Furthermore, we consider only an extract of the data available in scientific texts: the title and abstract, and the full text or its other sections such as the conclusion or introduction may contain signals on scientific originality which are excluded from our study due to this data limitation. 
We also only considered a relatively simple generalized linear model for the prediction of citations by DSI, modelling with a hurdle model or two part model---modelling firstly whether a paper is cited and then a second stage or model which predicts if a paper is cited, how many citations it will receive---may significantly improve the correlation.

SciBERT in particular is a relatively old model, given the pace of development in natural language processing since its release. The dataset that it was trained on contained papers exclusively from the Biomedical and Computer Science domains. Considering these two facts about SciBERT we note that there may be more powerful or appropriate models designed to process scientific texts which may outperform SciBERT as a model suited for predicting scientific originality. 

Furthermore, in \cite{Johnson2023}, the algorithm to compute DSI utilised \enquote{middle} (hidden) layers 6 and 7 of BERT. The choice of which layers to use, as discussed in \cite{Johnson2023}, was selected through a combination of justifications: from \cite{jawahar-etal-2019-bert} that early and middle layers are sensitive to syntactic and semantic information, and correlations with human originality rankings. This selection of layers may be a significant factor in the lack of predictive power observed in this paper, it may be the case that selection of other layers from SciBERT to retrieve the embeddings from may have proven more effective in capturing the desired sensitivity to scientific originality. Such a study is not feasible without ground truth data, which as mentioned is unavailable.

\section{Future Work}
We believe there is rich ground for quantitative analyses that implement, compare, and evaluate metrics for inferring scientific originality from text. 
This allows the strengths and weaknesses of different models to be compared and enables the identification of correlations between measures, helping to reveal potential redundancies among the models.
The ultimate aims of this effort are that better-informed judgments on scientific originality can be made by utilising the best-performing ensembles of models, and that these metrics can be adopted by the scientometric and research-policy communities with confidence.

To enable such future work, it would be useful to have a ground truth dataset of human measured originality scores for scientific papers, such as the studies in \cite{Johnson2023}. This will enable examining the hypothesis that DSI generalizes from short creative language texts to longer scientific texts, and furthermore correlates with scientific originality can be rigorously tested. Furthermore, such a dataset may allow for the exploration of hidden layer selection in the case of SciBERT for DSI computation, as described in Section \ref{sec:limitations}.

Future work may also include studying the sensitivity of DSI computation, and more generally the sensitivity of BERT- or other language model embeddings, to improperly formatted and out of distribution tokens in the context of scientometric analyses, such as those found in Quote \ref{quote:zhang-high-DSI} described in Section \ref{sec:limitations}. Another aspect of this is studying the overall scope of incorrect formatting of titles, abstracts and full-texts within large bibliometric databases.

We were initially surprised that SciBERT was not better performing in evaluating scientific originality. Thus, future research is needed to explore which models perform best in the task of semantically embedding the originality of scientific texts. Furthermore, model development such as fine tuning, or comparison of specialist models (such as SciBERT) against more modern general purpose models (such as large language models, or a wider range of smaller language models) is needed.

If, as mentioned in Section \ref{sec:limitations}, full texts were available for future study with DSI, we believe it would be interesting to study wider language models with larger maximum token input size. In particular, the incorporation of introduction, conclusions, related work and discussion sections (i.e. sections which contain descriptions of the science being considered in the paper). Furthermore, with respect to the limitation mentioned on the Gemini embeddings API only providing a single layer for consideration. Thus, we suggest self-hosted open source models such as Llama\footnote{\url{https://www.llama.com/}} may provide the flexibility to allow for considering multiple hidden layer embeddings.

Recent work by \cite{10.1162/qss_a_00109} has indicated that papers' citation impact should be considered in terms of their scope. They found that papers with deep citation impact typically focus on relatively narrow research areas, and those with broad citation impact typically cover a wider area of research. 
In our paper, when modelling citations after 5 years we considered only citation count rather than categorising the articles as those with a deep or broad citation impact. 
We therefore hypothesise that DSI may have more explanatory for articles with broader citation impact. We believe that articles with a broader research impact would have scientific terms stemming from wider research areas would likely be embedded more distantly,  and therefore measure a higher DSI.

\section{Conclusions}


In the current study, we introduce and explore a quantitative measure of originality--DSI--in quantifying the originality of scientific papers based on their titles and abstracts. Our work serves as a bridge between creativity and scientometric research, and highlights the opportunity of utilising metrics from creativity research in scientometric. 
While further research is needed to fully establish DSI’s contribution to scientometric research, our analyses—tracking its behaviour across scientific fields, publication years, author counts, and citation counts—provide compelling evidence of its feasibility. 
Overall, DSI captures a critical dimension of scientific papers—their originality—which plays a central role in shaping scientific contributions.

With reference to our three hypotheses laid out in Section \ref{sec:current-study}, we have demonstrated:

a) That BERT-DSI computed from combined titles and abstracts correlates with citation counts, even after controlling for other bibliometric variables, demonstrates its substantive explanatory power. 

b) That SciBERT-DSI scores do not exhibit stronger correlations with citation counts than those computed with BERT. In fact, we have demonstrated they do not correlate at all.

c) That BERT-DSI and SciBERT-DSI scores vary substantially across scientific domains, publication years, and author counts underscores the sensitivity of these metrics to key bibliometric dimensions.

\section{Data Availability}\label{sec:availability}
The code to compute both BERT- and SciBERT-DSI on GPU, as well as the code to compute the Gemini DSI can be found in the following repository: \url{https://github.com/jhculb/Scientometric-DSI}.

Unfortunately due to copyright restrictions sharing of the dataset alongside the paper is impossible, however we can provide a list of Web of Science identifiers and DOIs (where available in the dataset) which we do so here: \cite{culbert_2025_17778869}, these can be combined with the field mappings contained in Appendix Table \ref{tab:subjects-by-field} to reconstruct the dataset.

\section{Acknowledgements}
The authors acknowledge funding from the OMINO project (101086321), and Culbert and Mayr additionally acknowledge funding from the OpenBib project (16WIK2301B). 

\section{Conflicts of Interest}
Philipp Mayr, the co-author of this paper, has a conflict of interest because he serves on the editorial board of the journal Scientometrics.
\newpage
\bibliography{references.bib}

@article{https://doi.org/10.1111/j.1756-2171.2011.00140.x,
author = {Azoulay, Pierre and Graff Zivin, Joshua S. and Manso, Gustavo},
title = {Incentives and creativity: evidence from the academic life sciences},
journal = {The RAND Journal of Economics},
volume = {42},
number = {3},
pages = {527-554},
doi = {https://doi.org/10.1111/j.1756-2171.2011.00140.x},
year = {2011}
}

@Article{Beaty2023,
author={Beaty, Roger E.
and Kenett, Yoed N.},
title={Associative thinking at the core of creativity},
journal={Trends in Cognitive Sciences},
year={2023},
month={Jul},
day={01},
publisher={Elsevier},
volume={27},
number={7},
pages={671-683},
issn={1364-6613},
doi={10.1016/j.tics.2023.04.004},
url={https://doi.org/10.1016/j.tics.2023.04.004}
}

@inproceedings{beltagy-etal-2019-scibert,
    title = "{S}ci{BERT}: A Pretrained Language Model for Scientific Text",
    author = "Beltagy, Iz  and
      Lo, Kyle  and
      Cohan, Arman",
    editor = "Inui, Kentaro  and
      Jiang, Jing  and
      Ng, Vincent  and
      Wan, Xiaojun",
    booktitle = "Proceedings of the 2019 Conference on Empirical Methods in Natural Language Processing and the 9th International Joint Conference on Natural Language Processing (EMNLP-IJCNLP)",
    month = nov,
    year = "2019",
    address = "Hong Kong, China",
    publisher = "Association for Computational Linguistics",
    doi = "10.18653/v1/D19-1371",
    pages = "3615--3620"
}

@Article{Benedek2023,
author={Benedek, Mathias
and Beaty, Roger E.
and Schacter, Daniel L.
and Kenett, Yoed N.},
title={The role of memory in creative ideation},
journal={Nature Reviews Psychology},
year={2023},
month={Apr},
day={01},
volume={2},
number={4},
pages={246-257},
issn={2731-0574},
doi={10.1038/s44159-023-00158-z},
url={https://doi.org/10.1038/s44159-023-00158-z}
}

@article{Boudreau2016,
author = {Boudreau, Kevin J. and Guinan, Eva C. and Lakhani, Karim R. and Riedl, Christoph},
title = {Looking Across and Looking Beyond the Knowledge Frontier: Intellectual Distance, Novelty, and Resource Allocation in Science},
journal = {Management Science},
volume = {62},
number = {10},
pages = {2765-2783},
year = {2016},
doi = {10.1287/mnsc.2015.2285}
}

@misc{devlin2019bertpretrainingdeepbidirectional,
      title={BERT: Pre-training of Deep Bidirectional Transformers for Language Understanding}, 
      author={Jacob Devlin and Ming-Wei Chang and Kenton Lee and Kristina Toutanova},
      year={2019},
  archivePrefix={arXiv},
      doi={https://doi.org/10.48550/arXiv.1810.04805}, 
}

@Article{Johnson2023,
author={Johnson, Dan R.
and Kaufman, James C.
and Baker, Brendan S.
and Patterson, John D.
and Barbot, Baptiste
and Green, Adam E.
and van Hell, Janet
and Kennedy, Evan
and Sullivan, Grace F.
and Taylor, Christa L.
and Ward, Thomas
and Beaty, Roger E.},
title={Divergent semantic integration (DSI): Extracting creativity from narratives with distributional semantic modeling},
journal={Behavior Research Methods},
year={2023},
month={Oct},
day={01},
volume={55},
number={7},
pages={3726-3759},
issn={1554-3528},
doi={10.3758/s13428-022-01986-2}
}

@Article{Shibayama2020,
author={Shibayama, Sotaro
and Wang, Jian},
title={Measuring originality in science},
journal={Scientometrics},
year={2020},
month={Jan},
day={01},
volume={122},
number={1},
pages={409-427},
issn={1588-2861},
doi={10.1007/s11192-019-03263-0},
url={https://doi.org/10.1007/s11192-019-03263-0}
}

@article{TRAPIDO20151488,
title = {How novelty in knowledge earns recognition: The role of consistent identities},
journal = {Research Policy},
volume = {44},
number = {8},
pages = {1488-1500},
year = {2015},
issn = {0048-7333},
doi = {https://doi.org/10.1016/j.respol.2015.05.007},
url = {https://www.sciencedirect.com/science/article/pii/S0048733315000839},
author = {Denis Trapido}
}

@article{holyst024,
	title = {Protect our environment from information overload},
	issn = {2397-3374},
	url = {https://www.nature.com/articles/s41562-024-01833-8},
	doi = {10.1038/s41562-024-01833-8},
	journal = {Nature Human Behaviour},
	shortjournal = {Nat Hum Behav},
	author = {Ho\l{}yst, Janusz A. and Mayr, Philipp and Thelwall, Michael and Frommholz, Ingo and Havlin, Shlomo and Sela, Alon and Kenett, Yoed N. and Helic, Denis and Rehar, Aljosa and Ma\v{c}ek, Sebastijan R. and Kazienko, Przemys\l{}aw and Kajdanowicz, Tomasz and Biecek, Przemysław and Szymanski, Boleslaw K. and Sienkiewicz, Julian},
	date = {2024-02-07},
	year = {2024},
	langid = {english},
}

@INPROCEEDINGS{culbertISSI2025,
  title      = "Originality in scientific titles and abstracts can predict
                citation count",
  booktitle  = "International Conference on Scientometrics \& Informetrics",
  author     = "Culbert, Jack and Kenett, Yoed and Mayr, Philipp",
  month      =  jul,
  year       =  2025,
  conference = "20th International Conference on Scientometrics \& Informetrics",
  pages = {2283--2290},
  doi = "https://doi.org/10.51408/issi2025_106"
}

@article{ WOS:000620231200001,
Author = {Vicol, Ioana},
Title = {Multi-aged forest fragments in Atlantic France that are surrounded by
   meadows retain a richer epiphyte lichen flora},
Journal = {CRYPTOGAMIE MYCOLOGIE},
Year = {2020},
Volume = {41},
Number = {15},
Pages = {235-247},
Month = {DEC},

Publisher = {ADAC-CRYPTOGAMIE},
Address = {12 RUE DE BUFFON, 75005 PARIS, FRANCE},
Type = {Article},
Language = {English},
Affiliation = {Vicol, I (Corresponding Author), Romanian Acad, Inst Biol Bucharest, Dept Ecol Taxon \& Nat Conservat, 296 Splaiul Independentei St, Bucharest 060031, Romania.
   Vicol, Ioana, Romanian Acad, Inst Biol Bucharest, Dept Ecol Taxon \& Nat Conservat, 296 Splaiul Independentei St, Bucharest 060031, Romania.},
ISSN = {0181-1584},
EISSN = {1776-100X},
Keywords-Plus = {LAND-USE; DIVERSITY; SCALE; MANAGEMENT; ABUNDANCE; ECOLOGY; QUALITY;
   EUROPE; OAKS; TREE},
Research-Areas = {Mycology},
Web-of-Science-Categories  = {Mycology},
Author-Email = {ioana21vicol@gmail.com},
Affiliations = {Romanian Academy; Institute of Biology Bucharest},
ResearcherID-Numbers = {Vicol, Ioana/AAS-9474-2020},
Funding-Acknowledgement = {project ``Integrated European Long-Term Ecosystem \& Socio-ecological
   Research Infrastructure - eLTER, H2020 European Research Council{''}
   {[}GA 654359]},
Funding-Text = {The author is grateful to Vicol Constan.a and Dr Sandu Cristina for
   initial financial support to perform this study. I am indebted to Pat
   Wolseley (Natural History Museum, London, United Kingdom) and an
   anonymous reviewer for critical comments on the manuscript.
   Additionally, I thank the staff of the Centre national de la Recherche
   scientifique (France) especially to Bretagnolle Vincent for their
   assistance. American Journal Experts significantly contributed to
   improving the English language in the manuscript. The study was funded
   by the project ``Integrated European Long-Term Ecosystem \&
   Socio-ecological Research Infrastructure - eLTER, H2020 European
   Research Council{''}, (grant number GA 654359, under H2020).},
Number-of-Cited-References = {56},
Times-Cited = {2},
Usage-Count-Last-180-days = {0},
Usage-Count-Since-2013 = {5},
Journal-ISO = {Cryptogam. Mycol.},
Doc-Delivery-Number = {QK2SD},
Web-of-Science-Index = {Science Citation Index Expanded (SCI-EXPANDED)},
Unique-ID = {WOS:000620231200001},
DA = {2025-11-27},
doi="https://doi.org/10.5252/cryptogamie-mycologie2020v41a15"
}

@article{ WOS:000552521900001,
Author = {Zhang, Zhi-Feng and Zhou, Shi-Yue and Eurwilaichitr, Lily and
   Ingsriswang, Supawadee and Raza, Mubashar and Chen, Qian and Zhao, Peng
   and Liu, Fang and Cai, Lei},
Title = {Culturable mycobiota from Karst caves in China II, with descriptions of
   33 new species},
Journal = {FUNGAL DIVERSITY},
Year = {2021},
Volume = {106},
Number = {1, SI},
Pages = {29-136},
Month = {JAN},
Publisher = {SPRINGER},
Address = {ONE NEW YORK PLAZA, SUITE 4600, NEW YORK, NY, UNITED STATES},
Type = {Article},
Language = {English},
Affiliation = {Cai, L (Corresponding Author), Chinese Acad Sci, Inst Microbiol, State Key Lab Mycol, Beijing 100101, Peoples R China.
   Cai, L (Corresponding Author), Univ Chinese Acad Sci, Beijing 100049, Peoples R China.
   Zhang, Zhi-Feng; Zhou, Shi-Yue; Raza, Mubashar; Chen, Qian; Zhao, Peng; Liu, Fang; Cai, Lei, Chinese Acad Sci, Inst Microbiol, State Key Lab Mycol, Beijing 100101, Peoples R China.
   Zhang, Zhi-Feng, Shenzhen Univ, Inst Adv Study, Shenzhen 518060, Peoples R China.
   Zhang, Zhi-Feng; Zhou, Shi-Yue; Raza, Mubashar; Cai, Lei, Univ Chinese Acad Sci, Beijing 100049, Peoples R China.
   Eurwilaichitr, Lily; Ingsriswang, Supawadee, Natl Ctr Genet Engn \& Biotechnol, Thailand Bioresource Res Ctr, Bangkok, Thailand.},
DOI = {10.1007/s13225-020-00453-7},
EarlyAccessDate = {JUL 2020},
ISSN = {1560-2745},
EISSN = {1878-9129},
Keywords-Plus = {WHITE-NOSE SYNDROME; NATIONAL-PARK; SP-NOV; AIRBORNE FUNGI; ROCK
   SURFACES; INSECTICIDAL CYCLODEPSIPEPTIDES; HISTOPLASMA-CAPSULATUM;
   MICROBIAL COMMUNITIES; MULTIGENE PHYLOGENY; BEAUVERIA-BASSIANA},
Research-Areas = {Mycology},
Web-of-Science-Categories  = {Mycology},
Author-Email = {cail@im.ac.cn},
Affiliations = {Chinese Academy of Sciences; Institute of Microbiology, CAS; Shenzhen
   University; Chinese Academy of Sciences; University of Chinese Academy
   of Sciences, CAS; National Science \& Technology Development Agency -
   Thailand; National Center Genetic Engineering \& Biotechnology (BIOTEC)},
ResearcherID-Numbers = {Zhang, Zhi-Feng/AAW-1707-2020
   Cai, Lei/W-4996-2017
   Raza, Mubashar/AAO-1659-2020
   Eurwilaichitr, Lily/E-1032-2011
   Liu, Fang/HNJ-4139-2023
   Chen, Qian/JDC-8385-2023},
ORCID-Numbers = {Zhang, Zhi-Feng/0000-0002-9001-9850
   Raza, Mubashar/0000-0001-7665-1457
   CAI, LEI/0000-0002-8131-7274
   },
Funding-Acknowledgement = {NSFC {[}31725001]; Science and Technology Partnership Program, MOST
   {[}KY201701011]; Gansu Foundation of Ecological Conservation Remediation
   {[}2018-20]; Gansu Foundation of Inducing Scientific Innovation for
   Development {[}2017zx-10]},
Funding-Text = {This study was financially supported by NSFC (31725001), the Science and
   Technology Partnership Program, MOST (KY201701011), Gansu Foundation of
   Ecological Conservation \& Remediation (No. 2018-20) and Gansu
   Foundation of Inducing Scientific Innovation for Development (No.
   2017zx-10). Prof. Yuan-Hai Zhang in Institute of Karst Geology, Chinese
   Academy of Geological Sciences is thanked for providing caves'
   information in Southwest China. Dr. Ya-Li Xi in Gansu Engineering
   Laboratory of Applied Mycology, Hexi University is thanked for help with
   sample collection. We also thank other members who provided technical
   support, valuable and constructive suggestions in our lab.},
Number-of-Cited-References = {217},
Times-Cited = {98},
Usage-Count-Last-180-days = {7},
Usage-Count-Since-2013 = {101},
Journal-ISO = {Fungal Divers.},
Doc-Delivery-Number = {RM3IG},
Web-of-Science-Index = {Science Citation Index Expanded (SCI-EXPANDED)},
Unique-ID = {WOS:000552521900001},
DA = {2025-11-27},
}

@inproceedings{poleksic2023-ceurws,
    author = "Poleksi\'{c}, Andrija and Martinči\'{c}-Ip\v{s}i\'{c}, Sanda",
    title = "Effects of Pretraining Corpora on Scientific Relation Extraction Using BERT and SciBERT",
    note = {\url{https://ceur-ws.org/Vol-3510/paper_nlp_3.pdf}},
    crossref = {SEMANTiCS-Workshops2023}
}

@proceedings{SEMANTiCS-Workshops2023,
booktitle = {SEMANTiCS Workshops Proceedings Compound Volume 2023},
year = 2023,
editor = {D'Souza, Jennifer and Rula, Anisa and Chaves-Fraga, David and Sadeghi, Mersedeh and Hosseini Sohi, Shahrom and And\'{r}es Rojas, Juli\'{a}n and Colpaert, Pieter and Vakaj, Edlira and Tiwari, Sanju and Vahdati, Sahar and Lisa Gentile, Anna},

number = 3510,
series = {CEUR Workshop Proceedings},
address = {Aachen},
issn = {1613-0073},
url = {https://ceur-ws.org/Vol-3510/},
venue = {Leipzig, Germany},
eventdate = {2023-09-20},
title = {Joint Workshop Proceedings of the 5th International Workshop on A Semantic Data Space For Transport (Sem4Tra) and 2nd NLP4KGC: Natural Language Processing for Knowledge Graph Construction co-located with the 19th International Conference on Semantic Systems (SEMANTiCS 2023)}
}

@InProceedings{ming2020,
author="Jiang, Ming
and D'Souza, Jennifer
and Auer, S{\"o}ren
and Downie, J. Stephen",
editor="Ishita, Emi
and Pang, Natalie Lee San
and Zhou, Lihong",
title="Improving Scholarly Knowledge Representation: Evaluating BERT-Based Models for Scientific Relation Classification",
booktitle="Digital Libraries at Times of Massive Societal Transition",
year="2020",
publisher="Springer International Publishing",
address="Cham",
pages="3--19",
isbn="978-3-030-64452-9",
doi = "https://doi.org/10.1007/978-3-030-64452-9_1"
}

@InProceedings{guangyuan2021,
author="Piao, Guangyuan",
editor="Gupta, Manish
and Ramakrishnan, Ganesh",
title="Scholarly Text Classification with Sentence BERT and Entity Embeddings",
booktitle="Trends and Applications in Knowledge Discovery and Data Mining",
year="2021",
publisher="Springer International Publishing",
address="Cham",
pages="79--87",
isbn="978-3-030-75015-2",
doi = "https://doi.org/10.1007/978-3-030-75015-2_8"
}

@article{shibayama2021,
    doi = {10.1371/journal.pone.0254034},
    author = {Shibayama, Sotaro AND Yin, Deyun AND Matsumoto, Kuniko},
    journal = {PLOS ONE},
    publisher = {Public Library of Science},
    title = {Measuring novelty in science with word embedding},
    year = {2021},
    month = {07},
    volume = {16},
    url = {https://doi.org/10.1371/journal.pone.0254034},
    pages = {1-16},
    number = {7},

}

@article{ORGANISCIAK2023101356,
title = {Beyond semantic distance: Automated scoring of divergent thinking greatly improves with large language models},
journal = {Thinking Skills and Creativity},
volume = {49},
pages = {101356},
year = {2023},
issn = {1871-1871},
doi = {https://doi.org/10.1016/j.tsc.2023.101356},
url = {https://www.sciencedirect.com/science/article/pii/S1871187123001256},
author = {Peter Organisciak and Selcuk Acar and Denis Dumas and Kelly Berthiaume}
}

@ARTICLE{Beaty2021-bu,
  title     = "Automating creativity assessment with {SemDis}: An open platform
               for computing semantic distance",
  author    = "Beaty, Roger E and Johnson, Dan R",
  journal   = "Behav. Res. Methods",
  publisher = "Springer Science and Business Media LLC",
  volume    =  53,
  number    =  2,
  pages     = "757--780",
  month     =  apr,
  year      =  2021,
  copyright = "https://creativecommons.org/licenses/by/4.0",
  language  = "en",
  doi       = "https://doi.org/10.3758/s13428-020-01453-w"
}

@ARTICLE{Kenett2019-ub,
  title     = "What can quantitative measures of semantic distance tell us
               about creativity?",
  author    = "Kenett, Yoed N",
  journal   = "Current Opinion in Behaviour Sciences",
  publisher = "Elsevier BV",
  volume    =  27,
  pages     = "11--16",
  month     =  jun,
  year      =  2019,
  language  = "en",
  doi       = "https://doi.org/10.1016/j.cobeha.2018.08.010"
}

@article{Kenett2024-jv,
author = {Yoed N. Kenett},
title = {The Role of Knowledge in Creative Thinking},
journal = {Creativity Research Journal},
volume = {37},
number = {2},
pages = {242--249},
year = {2025},
publisher = {Routledge},
doi = {10.1080/10400419.2024.2322858},
URL = { 
        https://doi.org/10.1080/10400419.2024.2322858
}
}

@ARTICLE{Patterson2025-wb,
  title     = "{CAP}: The creativity assessment platform for online testing and
               automated scoring",
  author    = "Patterson, John D and Pronchick, Jimmy and Panchanadikar, Ruchi
               and Fuge, Mark and van Hell, Janet G and Miller, Scarlett R and
               Johnson, Dan R and Beaty, Roger E",
  publisher = "Springer",
  year      =  2025,
  doi       = "https://doi.org/https://doi.org/10.3929/ethz-c-000786692"
}

@ARTICLE{Runco2012-ou,
  title     = "The standard definition of creativity",
  author    = "Runco, Mark A and Jaeger, Garrett J",
  journal   = "Creat. Res. J.",
  publisher = "Informa UK Limited",
  volume    =  24,
  number    =  1,
  pages     = "92--96",
  month     =  jan,
  year      =  2012,
  language  = "en",
  doi = "https://psycnet.apa.org/doi/10.1080/10400419.2012.650092"
}

@article{Green02072024,
author = {Adam E. Green and Roger E. Beaty and Yoed N. Kenett and James C. Kaufman},
title = {The Process Definition of Creativity},
journal = {Creativity Research Journal},
volume = {36},
number = {3},
pages = {544--572},
year = {2024},
publisher = {Routledge},
doi = {10.1080/10400419.2023.2254573},
URL = {
        https://doi.org/10.1080/10400419.2023.2254573
}
}

@ARTICLE{Mikolov2013-en,
  title        = "Efficient estimation of word representations in vector space",
  author       = "Mikolov, Tomas and Chen, Kai and Corrado, Greg and Dean,
                  Jeffrey",
  year         =  2013,
  archivePrefix={arXiv},
  doi           = "https://doi.org/10.48550/arXiv.1301.3781"
}

@ARTICLE{Jiang2023-uk,
  title        = "{NeRT}: Implicit neural representations for general
                  unsupervised turbulence mitigation",
  author       = "Jiang, Weiyun and Liu, Yuhao and Boominathan, Vivek and
                  Veeraraghavan, Ashok",
  year         =  2023,
  archivePrefix={arXiv},
  doi = "https://doi.org/10.48550/arXiv.2308.00622"
}

@ARTICLE{Gray2019-zf,
  title     = "``Forward flow'': A new measure to quantify free thought and
               predict creativity",
  author    = "Gray, Kurt and Anderson, Stephen and Chen, Eric Evan and Kelly,
               John Michael and Christian, Michael S and Patrick, John and
               Huang, Laura and Kenett, Yoed N and Lewis, Kevin",
  journal   = "Am. Psychol.",
  publisher = "American Psychological Association (APA)",
  volume    =  74,
  number    =  5,
  pages     = "539--554",
  month     =  jul,
  year      =  2019,
  language  = "en"
}

@ARTICLE{Campidelli2026-sb,
  title     = "Creativity, the fountain of youth: Association between
               creativity and semantic memory networks across the lifespan",
  author    = "Campidelli, Lorenzo and Domanti, Umberto and Fusi, Giulia and
               Kenett, Yoed N and Agnoli, Sergio",
  journal   = "Cognition",
  publisher = "Elsevier BV",
  volume    =  266,
  number    =  106318,
  pages     = "106318",
  month     =  jan,
  year      =  2026,
  copyright = "http://creativecommons.org/licenses/by-nc-nd/4.0/",
  language  = "en",
  doi       = "https://doi.org/10.1016/j.cognition.2025.106318"
}

@inproceedings{paige-etal-2024-training,
    title = "Training {LLM}s to Recognize Hedges in Dialogues about Roadrunner Cartoons",
    author = "Paige, Amie  and
      Soubki, Adil  and
      Murzaku, John  and
      Rambow, Owen  and
      Brennan, Susan E.",
    editor = "Kawahara, Tatsuya  and
      Demberg, Vera  and
      Ultes, Stefan  and
      Inoue, Koji  and
      Mehri, Shikib  and
      Howcroft, David  and
      Komatani, Kazunori",
    booktitle = "Proceedings of the 25th Annual Meeting of the Special Interest Group on Discourse and Dialogue",
    month = sep,
    year = "2024",
    address = "Kyoto, Japan",
    publisher = "Association for Computational Linguistics",
    doi = "10.18653/v1/2024.sigdial-1.18",
    pages = "204--215"
}

@article{BAI2019407,
title = {Predicting the citations of scholarly paper},
journal = {Journal of Informetrics},
volume = {13},
number = {1},
pages = {407-418},
year = {2019},
issn = {1751-1577},
doi = {https://doi.org/10.1016/j.joi.2019.01.010},
url = {https://www.sciencedirect.com/science/article/pii/S1751157718301767},
author = {Xiaomei Bai and Fuli Zhang and Ivan Lee}
}

@article{ROBSON201694,
title = {Can we predict citation counts of environmental modelling papers? Fourteen bibliographic and categorical variables predict less than 30\% of the variability in citation counts},
journal = {Environmental Modelling \& Software},
volume = {75},
pages = {94-104},
year = {2016},
issn = {1364-8152},
doi = {https://doi.org/10.1016/j.envsoft.2015.10.007},
url = {https://www.sciencedirect.com/science/article/pii/S1364815215300657},
author = {Barbara J. Robson and A. {Mousqu{\`e}s}}
}

@article{10.1007/s11192-014-1279-6,
author = {Yu, Tian and Yu, Guang and Li, Peng-Yu and Wang, Liang},
title = {Citation impact prediction for scientific papers using stepwise regression analysis},
year = {2014},
issue_date = {November  2014},
publisher = {Springer-Verlag},
address = {Berlin, Heidelberg},
volume = {101},
number = {2},
issn = {0138-9130},
url = {https://doi.org/10.1007/s11192-014-1279-6},
doi = {10.1007/s11192-014-1279-6},
journal = {Scientometrics},
month = nov,
pages = {1233–1252},
numpages = {20}
}

@article{ZHAO2022101235,
title = {Utilizing citation network structure to predict paper citation counts: A Deep learning approach},
journal = {Journal of Informetrics},
volume = {16},
number = {1},
pages = {101235},
year = {2022},
issn = {1751-1577},
doi = {https://doi.org/10.1016/j.joi.2021.101235},
url = {https://www.sciencedirect.com/science/article/pii/S1751157721001061},
author = {Qihang Zhao and Xiaodong Feng}
}

@article{ABRAMO2025101725,
title = {Enhancing the prediction of publications’ long-term impact using early citations, readerships, and non-scientific factors},
journal = {Journal of Informetrics},
volume = {19},
number = {4},
pages = {101725},
year = {2025},
issn = {1751-1577},
doi = {https://doi.org/10.1016/j.joi.2025.101725},
url = {https://www.sciencedirect.com/science/article/pii/S1751157725000872},
author = {Giovanni Abramo and Tindaro Cicero and Ciriaco Andrea D’Angelo}
}

@article{10.1162/qss_a_00043,
    author = {Fraser, Nicholas and Momeni, Fakhri and Mayr, Philipp and Peters, Isabella},
    title = {The relationship between bioRxiv preprints, citations and altmetrics},
    journal = {Quantitative Science Studies},
    volume = {1},
    number = {2},
    pages = {618-638},
    year = {2020},
    month = {06},
    issn = {2641-3337},
    doi = {10.1162/qss_a_00043},
    url = {https://doi.org/10.1162/qss_a_00043}
}

@ARTICLE{Zhao2025-uz,
  title     = "A review on the novelty measurements of academic papers",
  author    = "Zhao, Yi and Zhang, Chengzhi",
  journal   = "Scientometrics",
  publisher = "Springer Science and Business Media LLC",
  volume    =  130,
  number    =  2,
  pages     = "727--753",
  month     =  feb,
  year      =  2025,
  copyright = "https://www.springernature.com/gp/researchers/text-and-data-mining",
  language  = "en",
  doi       = "https://doi.org/10.1007/s11192-025-05234-0"
}

@article{
doi:10.1126/science.1240474,
author = {Brian Uzzi  and Satyam Mukherjee  and Michael Stringer  and Ben Jones },
title = {Atypical Combinations and Scientific Impact},
journal = {Science},
volume = {342},
number = {6157},
pages = {468-472},
year = {2013},
doi = {10.1126/science.1240474}}

@article{WANG20171416,
title = {Bias against novelty in science: A cautionary tale for users of bibliometric indicators},
journal = {Research Policy},
volume = {46},
number = {8},
pages = {1416-1436},
year = {2017},
issn = {0048-7333},
doi = {https://doi.org/10.1016/j.respol.2017.06.006},
url = {https://www.sciencedirect.com/science/article/pii/S0048733317301038},
author = {Jian Wang and Reinhilde Veugelers and Paula Stephan}
}

@ARTICLE{Liu2022-vt,
  title     = "Pandemics are catalysts of scientific novelty: Evidence from
               {COVID-19}",
  author    = "Liu, Meijun and Bu, Yi and Chen, Chongyan and Xu, Jian and Li,
               Daifeng and Leng, Yan and Freeman, Richard B and Meyer, Eric T
               and Yoon, Wonjin and Sung, Mujeen and Jeong, Minbyul and Lee,
               Jinhyuk and Kang, Jaewoo and Min, Chao and Song, Min and Zhai,
               Yujia and Ding, Ying",
  journal   = "J. Assoc. Inf. Sci. Technol.",
  publisher = "Wiley",
  volume    =  73,
  number    =  8,
  pages     = "1065--1078",
  month     =  aug,
  year      =  2022,
  copyright = "http://onlinelibrary.wiley.com/termsAndConditions\#vor",
  language  = "en",
  doi       = "https://doi.org/10.1002/asi.24612"
}

@article{doi:10.1177/01655515231161133,
author = {Xuanmin Ruan and Weiyi Ao and Dongqing Lyu and Ying Cheng and Jiang Li},
title ={Effect of the topic-combination novelty on the disruption and impact of scientific articles: Evidence from PubMed},

journal = {Journal of Information Science},
volume = {51},
number = {5},
pages = {1033-1046},
year = {2025},
doi = {10.1177/01655515231161133},
}

@misc{priem2022openalexfullyopenindexscholarly,
      title={OpenAlex: A fully-open index of scholarly works, authors, venues, institutions, and concepts}, 
      author={Jason Priem and Heather Piwowar and Richard Orr},
      year={2022},
        archivePrefix={arXiv},
      doi={https://doi.org/10.48550/arXiv.2205.01833}, 
}

@article{chen2019,
  title={An Automatic Method for Extracting Innovative Ideas Based on the Scopus® Database},
  author={Lielei Chen, Hui Fang},
  journal={KO},
  volume={46},
  number={3},
  pages={171--186},
  doi = {10.5771/0943-7444-2019-3-171},
  year={2019}
}

@article{JEON2023101450,
title = {Measuring the novelty of scientific publications: A fastText and local outlier factor approach},
journal = {Journal of Informetrics},
volume = {17},
number = {4},
pages = {101450},
year = {2023},
issn = {1751-1577},
doi = {https://doi.org/10.1016/j.joi.2023.101450},
url = {https://www.sciencedirect.com/science/article/pii/S1751157723000755},
author = {Daeseong Jeon and Junyoup Lee and Joon Mo Ahn and Changyong Lee}
}

@article{WANG2024101587,
title = {An effective framework for measuring the novelty of scientific articles through integrated topic modeling and cloud model},
journal = {Journal of Informetrics},
volume = {18},
number = {4},
pages = {101587},
year = {2024},
issn = {1751-1577},
doi = {https://doi.org/10.1016/j.joi.2024.101587},
url = {https://www.sciencedirect.com/science/article/pii/S1751157724000993},
author = {Zhongyi Wang and Haoxuan Zhang and Jiangping Chen and Haihua Chen}
}

@inproceedings{neumann-etal-2019-scispacy,
    title = "{S}cispa{C}y: Fast and Robust Models for Biomedical Natural Language Processing",
    author = "Neumann, Mark  and
      King, Daniel  and
      Beltagy, Iz  and
      Ammar, Waleed",
    editor = "Demner-Fushman, Dina  and
      Cohen, Kevin Bretonnel  and
      Ananiadou, Sophia  and
      Tsujii, Junichi",
    booktitle = "Proceedings of the 18th BioNLP Workshop and Shared Task",
    month = aug,
    year = "2019",
    address = "Florence, Italy",
    publisher = "Association for Computational Linguistics",
    doi = "10.18653/v1/W19-5034",
    pages = "319--327"
}

@misc{vaswani2023attentionneed,
      title={Attention Is All You Need}, 
      author={Ashish Vaswani and Noam Shazeer and Niki Parmar and Jakob Uszkoreit and Llion Jones and Aidan N. Gomez and Lukasz Kaiser and Illia Polosukhin},
      year={2023},
  archivePrefix={arXiv},
      doi={https://doi.org/10.48550/arXiv.1706.03762}, 
}

@InProceedings{10.1007/978-3-031-65794-8_16,
author="Wolff, Benjamin
and Seidlmayer, Eva
and F{\"o}rstner, Konrad U.",
editor="Rehm, Georg
and Dietze, Stefan
and Schimmler, Sonja
and Kr{\"u}ger, Frank",
title="Enriched BERT Embeddings for Scholarly Publication Classification",
booktitle="Natural Scientific Language Processing and Research Knowledge Graphs",
year="2024",
publisher="Springer Nature Switzerland",
address="Cham",
pages="234--243",
isbn="978-3-031-65794-8",
doi = "10.1007/978-3-031-65794-8_16"
}

@inproceedings{10.1109/ICCV.2015.11,
author = {Zhu, Yukun and Kiros, Ryan and Zemel, Rich and Salakhutdinov, Ruslan and Urtasun, Raquel and Torralba, Antonio and Fidler, Sanja},
title = {Aligning Books and Movies: Towards Story-Like Visual Explanations by Watching Movies and Reading Books},
year = {2015},
isbn = {9781467383912},
publisher = {IEEE Computer Society},
address = {USA},
doi = {10.1109/ICCV.2015.11},
booktitle = {Proceedings of the 2015 IEEE International Conference on Computer Vision (ICCV)},
pages = {19–27},
numpages = {9},
series = {ICCV '15}
}

@article{https://doi.org/10.1002/jocb.658,
author = {Goecke, Benjamin and DiStefano, Paul V. and Aschauer, Wolfgang and Haim, Kurt and Beaty, Roger and Forthmann, Boris},
title = {Automated Scoring of Scientific Creativity in German},
journal = {The Journal of Creative Behavior},
volume = {58},
number = {3},
pages = {321-327},
doi = {https://doi.org/10.1002/jocb.658},
year = {2024}
}

@article{orwig2024,
author = {Orwig, William and Edenbaum, Emma R. and Greene, Joshua D. and Schacter, Daniel L.},
title = {The Language of Creativity: Evidence from Humans and Large Language Models},
journal = {The Journal of Creative Behavior},
volume = {58},
number = {1},
pages = {128-136},
doi = {https://doi.org/10.1002/jocb.636},
year = {2024}
}

@INPROCEEDINGS{11126815,
  author={Narayanan, Harini},
  booktitle={2025 IEEE 49th Annual Computers, Software, and Applications Conference (COMPSAC)}, 
  title={Quantifying creativity in {AI}-generated podcasts}, 
  year={2025},
  volume={},
  number={},
  pages={1613-1618},
  doi={10.1109/COMPSAC65507.2025.00217}}

@misc{taffa_leveraging_2023,
    title = {Leveraging {LLMs} in {Scholarly} {Knowledge} {Graph} {Question} {Answering}},
    doi = {10.48550/arXiv.2311.09841},
    publisher = {arXiv},
    author = {Taffa, Tilahun Abedissa and Usbeck, Ricardo},
    month = nov,
    year = {2023},
}

@misc{geminiteam2025geminifamilyhighlycapable,
      title={Gemini: A Family of Highly Capable Multimodal Models}, 
      author={{Gemini Team et al.}},
      year={2025},
      doi={https://doi.org/10.48550/arXiv.2312.11805}, 
}

@article{18bce95a-04fb-37a4-8c3e-29e9bbcae8ac,
 ISSN = {03067734, 17515823},
 URL = {http://www.jstor.org/stable/1403192},
 author = {Carlos M. Jarque and Anil K. Bera},
 journal = {International Statistical Review / Revue Internationale de Statistique},
 number = {2},
 pages = {163--172},
 publisher = {[Wiley, International Statistical Institute (ISI)]},
 title = {A Test for Normality of Observations and Regression Residuals},
 volume = {55},
 year = {1987},
 doi = "https://doi.org/10.2307/1403192"
}

@misc{loper2002nltknaturallanguagetoolkit,
      title={NLTK: The Natural Language Toolkit}, 
      author={Edward Loper and Steven Bird},
      year={2002},
  archivePrefix={arXiv},
      doi={https://doi.org/10.48550/arXiv.cs/0205028}, 
}

@article{ WOS:000340692800018,
Author = {Ayas, M. Bahadir and Sak, Ugur},
Title = {Objective measure of scientific creativity: Psychometric validity of the
   Creative Scientific Ability Test},
Journal = {Thinking Skills and Creativity},
Year = {2014},
Volume = {13},
Pages = {195-205},
DOI = {10.1016/j.tsc.2014.06.001},
ISSN = {1871-1871},
EISSN = {1878-0423},
ResearcherID-Numbers = {SAK, UGUR/AAT-7763-2020
   ayas, muhammet/O-4631-2014
   },
ORCID-Numbers = {SAK, Ugur/0000-0001-6312-5239},
Unique-ID = {WOS:000340692800018},
}

@article{LUO2022101282,
title = {Combination of research questions and methods: A new measurement of scientific novelty},
journal = {Journal of Informetrics},
volume = {16},
number = {2},
pages = {101282},
year = {2022},
issn = {1751-1577},
doi = {https://doi.org/10.1016/j.joi.2022.101282},
url = {https://www.sciencedirect.com/science/article/pii/S1751157722000347},
author = {Zhuoran Luo and Wei Lu and Jiangen He and Yuqi Wang}
}

@techreport{osti_6910294,
  author       = {Jordan, M I},
  title        = {Serial order: a parallel distributed processing approach. Technical report, June 1985-March 1986},
  institution  = {California Univ., San Diego, La Jolla (USA). Inst. for Cognitive Science},
  url          = {https://www.osti.gov/biblio/6910294},
  place        = {United States},
  year         = {1986},
  month        = {05}}

@article{10.1162/neco.1997.9.8.1735,
author = {Hochreiter, Sepp and Schmidhuber, J\"{u}rgen},
title = {Long Short-Term Memory},
year = {1997},
issue_date = {November 15, 1997},
publisher = {MIT Press},
address = {Cambridge, MA, USA},
volume = {9},
number = {8},
issn = {0899-7667},
url = {https://doi.org/10.1162/neco.1997.9.8.1735},
doi = {10.1162/neco.1997.9.8.1735},
journal = {Neural Comput.},
month = nov,
pages = {1735–1780},
numpages = {46}
}

@incollection{Rumelhart1986LearningIR,
title = {Learning Internal Representations by Error Propagation},
editor = {Allan Collins and Edward E. Smith},
booktitle = {Readings in Cognitive Science},
publisher = {Morgan Kaufmann},
pages = {399-421},
year = {1988},
isbn = {978-1-4832-1446-7},
doi = {https://doi.org/10.1016/B978-1-4832-1446-7.50035-2},
author = {D.E. Rumelhart and G.E. Hinton and R.J. Williams}
}

@misc{chung2014empiricalevaluationgatedrecurrent,
      title={Empirical Evaluation of Gated Recurrent Neural Networks on Sequence Modeling}, 
      author={Junyoung Chung and Caglar Gulcehre and KyungHyun Cho and Yoshua Bengio},
      year={2014},
  archivePrefix={arXiv},
      doi={https://doi.org/10.48550/arXiv.1412.3555}, 
}

@article{10.1162/QSS.a.20,
    author = {Schmidt, Marion and Rimmert, Christine and Stephen, Dimity and Lenke, Christopher and Donner, Paul and Gärtner, Simone and Taubert, Niels and Bausenwein, Thomas and Stahlschmidt, Stephan},
    title = {The data infrastructure of the German Kompetenznetzwerk Bibliometrie: An enabling intermediary between raw data and analysis},
    journal = {Quantitative Science Studies},
    volume = {6},
    pages = {1129-1146},
    year = {2025},
    month = {10},
    issn = {2641-3337},
    doi = {10.1162/QSS.a.20},
    url = {https://doi.org/10.1162/QSS.a.20}
}

@article{doi:10.1177/1745691619861372,
author = {Fritz G\"unther and Luca Rinaldi and Marco Marelli},
title ={Vector-Space Models of Semantic Representation From a Cognitive Perspective: A Discussion of Common Misconceptions},

journal = {Perspectives on Psychological Science},
volume = {14},
number = {6},
pages = {1006-1033},
year = {2019},
doi = {10.1177/1745691619861372},
    note ={PMID: 31505121}
}

@article{doi:10.1080/17470218.2014.988735,
author = {Paweł Mandera and Emmanuel Keuleers and Marc Brysbaert},
title ={How useful are corpus-based methods for extrapolating psycholinguistic variables?},

journal = {Quarterly Journal of Experimental Psychology},
volume = {68},
number = {8},
pages = {1623-1642},
year = {2015},
doi = {10.1080/17470218.2014.988735},
    note ={PMID: 25695623}
}

@article{MANDERA201757,
title = {Explaining human performance in psycholinguistic tasks with models of semantic similarity based on prediction and counting: A review and empirical validation},
journal = {Journal of Memory and Language},
volume = {92},
pages = {57-78},
year = {2017},
issn = {0749-596X},
doi = {https://doi.org/10.1016/j.jml.2016.04.001},
url = {https://www.sciencedirect.com/science/article/pii/S0749596X16300079},
author = {Paweł Mandera and Emmanuel Keuleers and Marc Brysbaert}
}

@ARTICLE{Adams2005-xp,
  title     = "Early citation counts correlate with accumulated impact",
  author    = "Adams, Jonathan",
  journal   = "Scientometrics",
  publisher = "Springer Science and Business Media LLC",
  volume    =  63,
  number    =  3,
  pages     = "567--581",
  month     =  jun,
  year      =  2005,
  language  = "en",
  doi       = "10.1007/s11192-005-0228-9",
}

@ARTICLE{Zhang2024-uc,
  title     = "Predicting citation impact of academic papers across research
               areas using multiple models and early citations",
  author    = "Zhang, Fang and Wu, Shengli",
  journal   = "Scientometrics",
  publisher = "Springer Science and Business Media LLC",
  volume    =  129,
  number    =  7,
  pages     = "4137--4166",
  month     =  jul,
  year      =  2024,
  copyright = "https://creativecommons.org/licenses/by/4.0",
  language  = "en",
    doi = "https://doi.org/10.1007/s11192-024-05086-0"
}

@article{10.1162/qss_a_00109,
    author = {Bu, Yi and Waltman, Ludo and Huang, Yong},
    title = {A multidimensional framework for characterizing the citation impact
                    of scientific publications},
    journal = {Quantitative Science Studies},
    volume = {2},
    number = {1},
    pages = {155-183},
    year = {2021},
    month = {04},
    issn = {2641-3337},
    doi = {10.1162/qss_a_00109},
    url = {https://doi.org/10.1162/qss_a_00109},
}

@article{10.1162/qss_a_00003,
    author = {Thelwall, Mike},
    title = {Large publishing consortia produce higher citation impact research but coauthor contributions are hard to evaluate},
    journal = {Quantitative Science Studies},
    volume = {1},
    number = {1},
    pages = {290-302},
    year = {2020},
    month = {02},
    issn = {2641-3337},
    doi = {10.1162/qss_a_00003},
}

@ARTICLE{Tahamtan2016-mr,
  title     = "Factors affecting number of citations: a comprehensive review of
               the literature",
  author    = "Tahamtan, Iman and Safipour Afshar, Askar and Ahamdzadeh,
               Khadijeh",
  journal   = "Scientometrics",
  publisher = "Springer Science and Business Media LLC",
  volume    =  107,
  number    =  3,
  pages     = "1195--1225",
  month     =  jun,
  year      =  2016,
  language  = "en",
    doi =   "https://doi.org/10.1007/s11192-016-1889-2"
}

@inproceedings{jawahar-etal-2019-bert,
    title = "What Does {BERT} Learn about the Structure of Language?",
    author = "Jawahar, Ganesh  and
      Sagot, Beno{\^i}t  and
      Seddah, Djam{\'e}",
    editor = "Korhonen, Anna  and
      Traum, David  and
      M{\`a}rquez, Llu{\'i}s",
    booktitle = "Proceedings of the 57th Annual Meeting of the Association for Computational Linguistics",
    month = jul,
    year = "2019",
    address = "Florence, Italy",
    publisher = "Association for Computational Linguistics",
    doi = "10.18653/v1/P19-1356",
    pages = "3651--3657",
}

@dataset{culbert_2025_17778869,
  author       = {Culbert, Jack},
  title        = {Dataset for Investigating the Originality of
                   Scientific Papers across Time and Domain: A
                   Quantitative Analysis
                  },
  month        = dec,
  year         = 2025,
  publisher    = {Zenodo},
  doi          = {10.5281/zenodo.17778869},
}
\newpage
\appendix
\section*{Appendix}
\section{CPU \& GPU Computation Comparison}\label{appx:gpu}

\begin{figure}[hb!]
    \centering
    \includegraphics[width=0.75\linewidth]{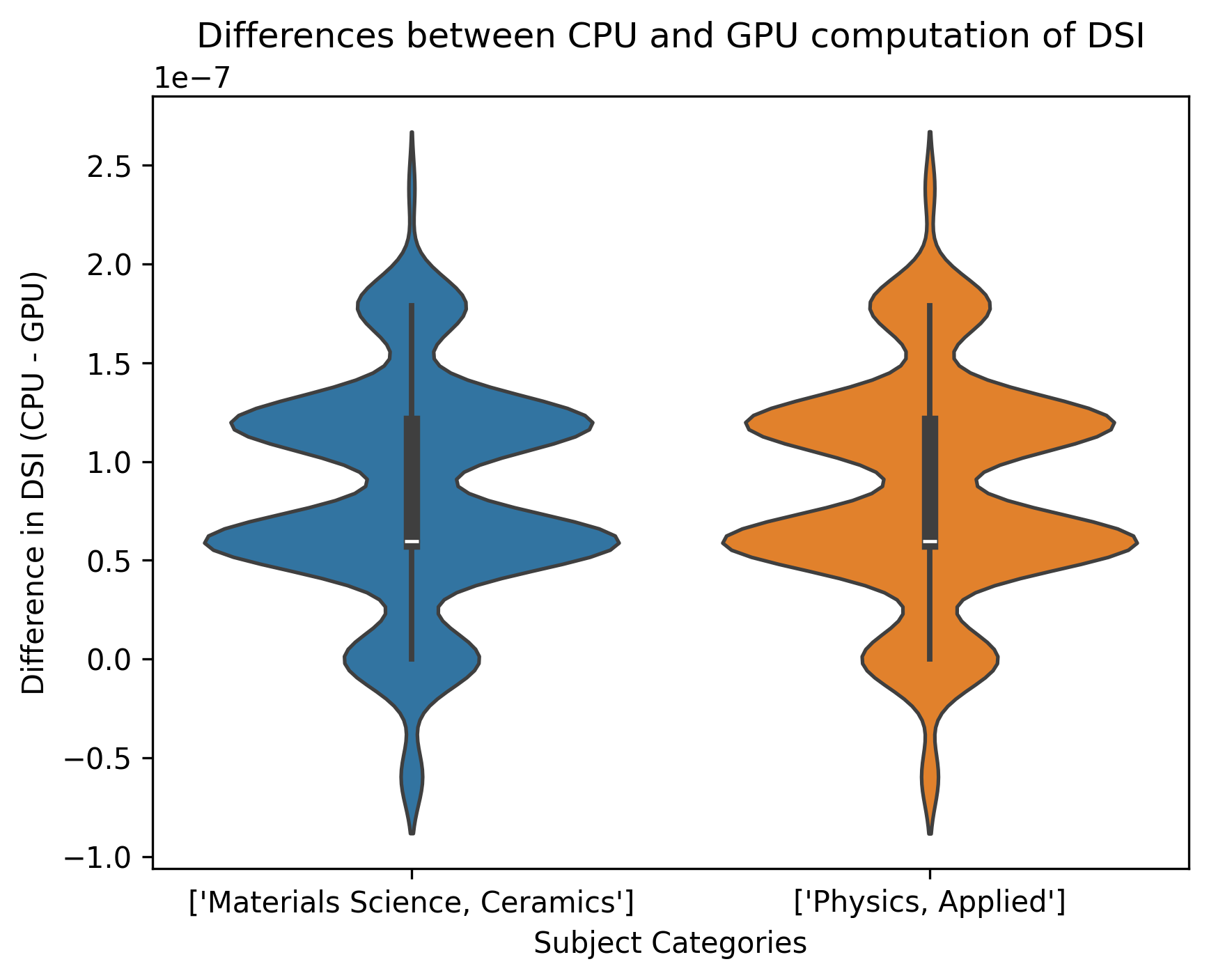}
    \caption{Violin plot of the differences between CPU and GPU computation of DSI using BERT}
    \label{fig:cpu-gpu-comparison}
\end{figure}

For rigour, we then compared the results from the GPU version of the code in two categories, Applied Physics and Ceramics, to verify that the results from the GPU code were equivalent to the results produced by the CPU version of the code. As seen in Figure \ref{fig:cpu-gpu-comparison}, the errors are on the scale of $1 \times 10^{-7}$, and are likely to do with the differences of floating point representations in CPU and GPU architecture, we therefore accepted these errors as insignificant enough to proceed with using GPU computation.

Our CPU computation was performed on a virtualised 32 core processor with 126GB of RAM available, the host system used a Intel Xeon Gold 6242 16 core, 32 thread CPU running at 2.8GHz with a maximum turbo boost frequency of 3.9 GHz. The GPU computation was performed on a n1-standard-32 T4 VM compute instance\footnote{\url{https://docs.cloud.google.com/compute/docs/gpus}} sourced from Google Cloud\footnote{\url{https://console.cloud.google.com}}, designed for general purpose GPU workloads. This instance contained 32vCPUs, 120GB of memory and had access to 1 NVIDIA T4 GPU.

\begin{table}[htbp!]
    \centering
    \begin{tabular}{l|cc|cc}
         \toprule
         & \multicolumn{2}{c|}{BERT-DSI CPU} & \multicolumn{2}{c}{BERT-DSI GPU} \\
         & Total & Per Abstract & Total & Per Abstract \\
         \midrule
         Applied Physics & 201614.8 & 227.813 & 165.7 & 0.187 \\
         Ceramics & 197035.1 & 222.639 & 166.5 & 0.188 \\
         \bottomrule
    \end{tabular}
    \caption{Seconds taken to compute BERT-DSI for two Primary Subjects for CPU and GPU computation, in total for each subject and per abstract}
    \label{tab:computation-times}
\end{table}

We note that in designing the GPU code it was observed that a significant part of the improvement in computation time through using GPU was the offloading of the cosine comparison function to the GPU. We did not complete the comparison for SciBERT in Table \ref{tab:computation-times}, as time and cost restrictions limited our study, but we observed a similar CPU and GPU computation times.

\section{Subject Selection and Field Classification}\label{appx:subjects}
In Table \ref{tab:subjects-by-field} we list the subjects used in the study and categorise them by Field of Research. As mentioned before the notable features are that Social Sciences has 19 categories compared to 20 in the other categories, barring Multidisciplinary Sciences which is a field with a single category.
\begin{table}[htbp]
\centering
\begin{tabular}{lp{9cm}}
\toprule
\textbf{Field of Research} & \textbf{Primary Subjects} \\
\midrule
Life Sciences \& Biomedicine & `Agriculture, Multidisciplinary', `Biophysics', `Cardiac \& Cardiovascular Systems', `Cell Biology', `Clinical Neurology', `Ecology', `Entomology', `Forestry', `Genetics \& Heredity', `Horticulture', `Immunology', `Medicine, Research \& Experimental', `Mycology', `Neurosciences', `Ornithology', `Pathology', `Pediatrics', `Pharmacology \& Pharmacy', `Physiology', `Urology \& Nephrology' \\
Multidisciplinary Sciences & `Multidisciplinary Sciences' \\
Physical Sciences & `Astronomy \& Astrophysics', `Chemistry, Analytical', `Chemistry, Inorganic \& Nuclear', `Chemistry, Medicinal', `Chemistry, Multidisciplinary', `Chemistry, Physical', `Electrochemistry', `Geochemistry \& Geophysics', `Geology', `Mathematics, Applied', `Meteorology \& Atmospheric Sciences', `Physics, Applied', `Physics, Atomic, Molecular \& Chemical', `Physics, Condensed Matter', `Physics, Fluids \& Plasmas', `Physics, Mathematical', `Physics, Multidisciplinary', `Physics, Nuclear', `Physics, Particles \& Fields', `Thermodynamics' \\
Social Sciences & `Business', `Economics', `Education \& Educational Research', `Education, Scientific Disciplines', `Geography', `Geography, Physical', `International Relations', `Management', `Political Science', `Psychology', `Psychology, Biological', `Psychology, Clinical', `Psychology, Developmental', `Psychology, Educational', `Psychology, Experimental', `Psychology, Multidisciplinary', `Psychology, Social', `Social Sciences, Interdisciplinary', `Sociology' \\
Technology & `Acoustics', `Computer Science, Artificial Intelligence', `Computer Science, Information Systems', `Computer Science, Interdisciplinary Applications', `Construction \& Building Technology', `Energy \& Fuels', `Engineering, Biomedical', `Engineering, Chemical', `Engineering, Electrical \& Electronic', `Engineering, Environmental', `Engineering, Mechanical', `Engineering, Multidisciplinary', `Information Science \& Library Science', `Instruments \& Instrumentation', `Materials Science, Ceramics', `Materials Science, Multidisciplinary', `Materials Science, Paper \& Wood', `Mechanics', `Nanoscience \& Nanotechnology', `Nuclear Science \& Technology' \\
\bottomrule
\end{tabular}
\caption{Primary Subjects by Field of Research}
\label{tab:subjects-by-field}
\end{table}
\section{LLM Embedding Model Exploration}\label{appx:gemini}
We also explored a further hypothesis in this paper:\\

Our hypothesis is that DSI computed with a Large Language Model (LLM) as the embedding model would have even fewer out-of-distribution tokens in this context, having been trained on scientific works, and may produce a stronger correlation to citation count than a BERT based embedding model.\\

To computing DSI with a LLM, we chose a Google Gemini embedding model \citep{geminiteam2025geminifamilyhighlycapable}, in particular the gemini-embedding-001 model. This requires some adjustment to the formula given in Section \ref{sec:DSI-description}, as Google Gemini provides access to only a single output vector. As such we simply calculate the cosine distance of the resulting output vectors returned for each sentence. This results in the following: for a given text $T$ represented as an ordered list of length $n>2$ containing sentences $s_{i}$, and the embedding vector from the Gemini model defined as $GEMINI(s_{i})=\beta\{(i)\}$:
 \begin{equation*}
    DSI([s_{1},s_{2},\ldots,s_{n}])=\sum_{1 \leq i < j \leq n}\frac{1-\frac{\beta_{i}\cdot\beta_{j}}{\|\beta_{i}\|\cdot\|\beta_{j}\|}}{2n} 
 \end{equation*} \label{eq:DSI-gemini}

\begin{figure}[hb!]
    \centering
    \includegraphics[width=1\linewidth]{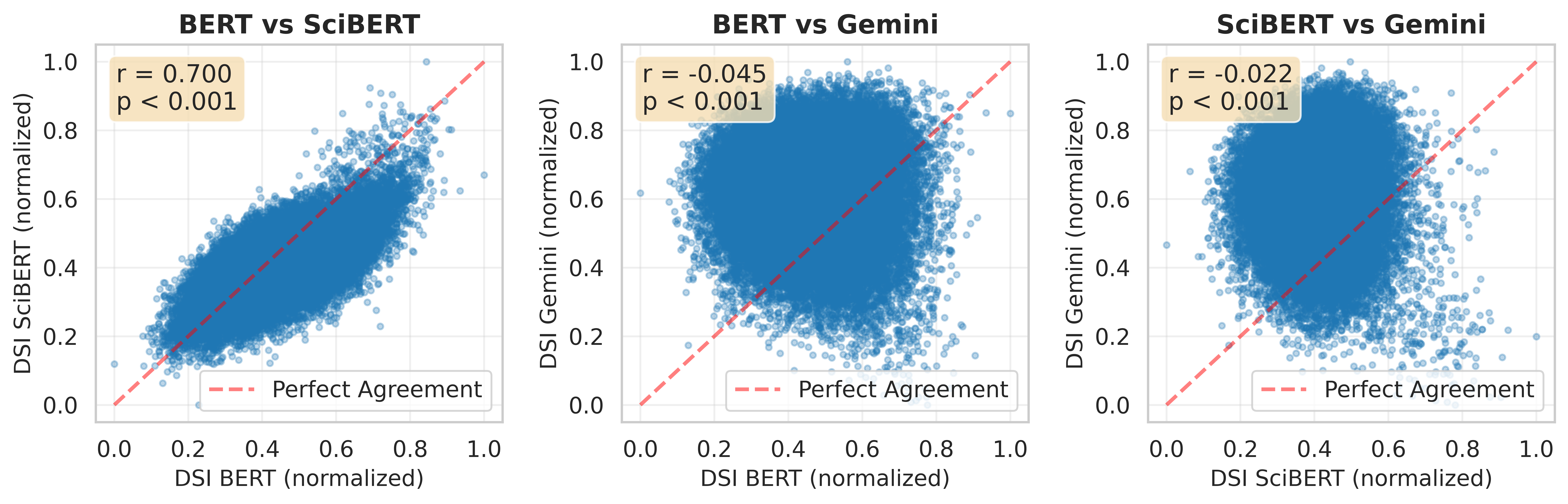}
    \caption{Scatter diagram of normalised DSI values when computed with different embedding models, Pearson Correlation r and corresponding p values given}
    \label{fig:gemini-pairwise}
\end{figure}

However, we found that the resulting Gemini-DSI numbers were not significantly correlated with BERT- nor SciBERT-DSI, even after normalisation, as demonstrated in Figure \ref{fig:gemini-pairwise}. \textcolor{black}{This implied that the signal identified in \cite{Johnson2023} was not present in the computation using Gemini}. Due to this finding \textcolor{black}{and the inference}, we prioritized work on the other two embedding models.

\section{Pairwise Comparison of BERT-DSI and SciBERT-DSI}
\begin{figure}[htbp]
    \centering
    \includegraphics[width=1\linewidth]{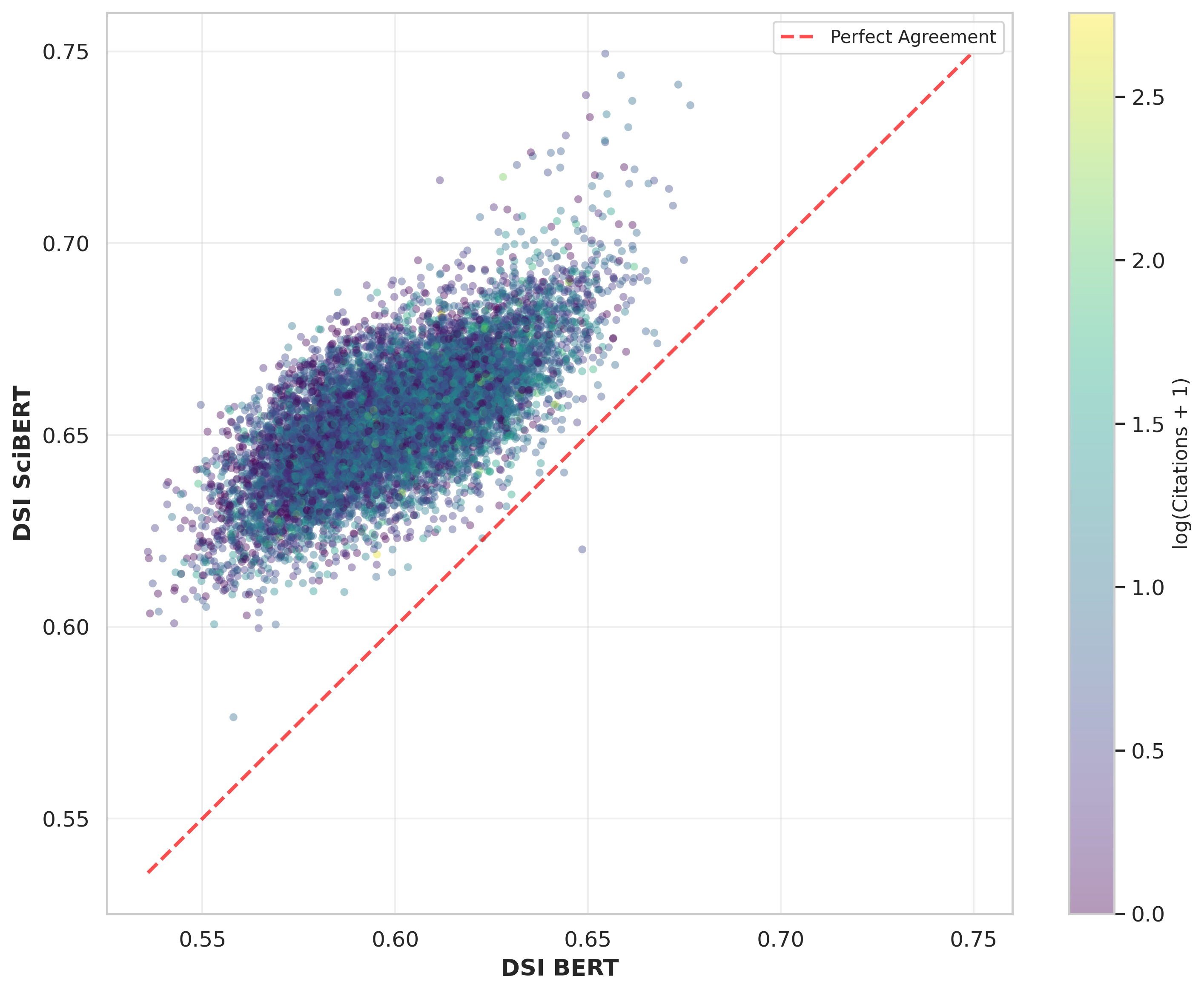}
    \caption{Pairwise scatter plot comparison of BERT-DSI and SCIBERT-DSI, with logarithm of citation count after 5 years + 1 as colour index}
    \label{fig:pairwise-BERT-SciBERT}
\end{figure}

In Figure \ref{fig:pairwise-BERT-SciBERT}, we plot the BERT-DSI and SciBERT-DSI for every paper in the dataset with the log of citations after 5 years plus + 1 as the colour. We observe a positive correlation between the two embedding models DSI scores, with a higher value for SciBERT-DSI than BERT-DSI across all but two papers. 
\end{document}